\documentclass[aps, prd,twocolumn,superscriptaddress,nofootinbib,preprintnumbers]{revtex4-1}
\def\araa{ARA\&A}%
\def\apj{ApJ}%
\def\apjs{ApJS}%
\def\aap{A\&A}%
\def\aapr{A\&A~Rev.}%
\def\mnras{MNRAS}%
\def\prd{Phys.~Rev.~D}%
\def\nat{Nature}%
\def\physrep{Phys.~Rep.}%
\include{notations}
\usepackage{slashed}
\usepackage{graphicx}
\graphicspath{{./figure/}}
\usepackage{grffile}
\usepackage[normalem]{ulem}
\usepackage{dcolumn}
\usepackage{xcolor}
\usepackage{booktabs}
\usepackage{amsmath}
\usepackage{amssymb}
\usepackage{slashed}
\usepackage{todonotes}
\usepackage{comment}
\usepackage{float}
\usepackage{natbib}
\usepackage{hyperref}
\usepackage{mathtools}

\newcommand{\beq}{\begin{equation}}
\newcommand{\eeq}{\end{equation}}
\newcommand{\beqa}{\begin{eqnarray}}
\newcommand{\eeqa}{\end{eqnarray}}

\usepackage{multirow}
\usepackage{arydshln}
\usepackage[T1]{fontenc}
\usepackage{ae,aecompl}
\usepackage{lineno}
\usepackage{eso-pic}% http://ctan.org/pkg/eso-pic
\usepackage{txfonts}

\newcommand{\redmagic}{\texttt{redMaGiC}\,}
\newcommand{\mice}{\texttt{MICE}\,}
\newcommand{\buzzard}{\texttt{Buzzard}\,}

\defcitealias{Battaglia:2012}{B12}
\defcitealias{Battaglia:2010}{B10}
\defcitealias{Vikram:2017}{V17}

\usepackage{lineno}
\begin{document}
\title[Constraining feedback with Galaxy-$y$ cross-correlations]{\boldmath Constraints on the redshift evolution of astrophysical feedback with Sunyaev-Zel'dovich effect cross-correlations}

% Author list file generated with: mkauthlist 1.2.4 
% mkauthlist -f -j prd DES-2019-0430_author_list_final.csv DES-2019-0430_author_list_final_arxiv.tex 

\author{S.~Pandey}
\affiliation{Department of Physics and Astronomy, University of Pennsylvania, Philadelphia, PA 19104, USA}
\author{E.~J.~Baxter}
\affiliation{Department of Physics and Astronomy, University of Pennsylvania, Philadelphia, PA 19104, USA}
\author{Z.~Xu}
\affiliation{Department of Physics and Astronomy, University of Pennsylvania, Philadelphia, PA 19104, USA}
\author{J.~Orlowski-Scherer}
\affiliation{Department of Physics and Astronomy, University of Pennsylvania, Philadelphia, PA 19104, USA}
\author{N.~Zhu}
\affiliation{Department of Physics and Astronomy, University of Pennsylvania, Philadelphia, PA 19104, USA}
\author{A.~Lidz}
\affiliation{Department of Physics and Astronomy, University of Pennsylvania, Philadelphia, PA 19104, USA}
\author{J.~Aguirre}
\affiliation{Department of Physics and Astronomy, University of Pennsylvania, Philadelphia, PA 19104, USA}
\author{J.~DeRose}
\affiliation{Department of Physics, Stanford University, 382 Via Pueblo Mall, Stanford, CA 94305, USA}
\affiliation{Kavli Institute for Particle Astrophysics \& Cosmology, P. O. Box 2450, Stanford University, Stanford, CA 94305, USA}
\author{M.~Devlin}
\affiliation{Department of Physics and Astronomy, University of Pennsylvania, Philadelphia, PA 19104, USA}
\author{J.~C.~Hill}
\affiliation{School of Natural Sciences, Institute for Advanced Study, Princeton, NJ, USA 08540}
\affiliation{Center for Computational Astrophysics, Flatiron Institute, New York, NY, USA 10003}
\author{B.~Jain}
\affiliation{Department of Physics and Astronomy, University of Pennsylvania, Philadelphia, PA 19104, USA}
\author{R.~K.~Sheth}
\affiliation{Department of Physics and Astronomy, University of Pennsylvania, Philadelphia, PA 19104, USA}
\author{S.~Avila}
\affiliation{Instituto de Fisica Teorica UAM/CSIC, Universidad Autonoma de Madrid, 28049 Madrid, Spain}
\author{E.~Bertin}
\affiliation{CNRS, UMR 7095, Institut d'Astrophysique de Paris, F-75014, Paris, France}
\affiliation{Sorbonne Universit\'es, UPMC Univ Paris 06, UMR 7095, Institut d'Astrophysique de Paris, F-75014, Paris, France}
\author{D.~Brooks}
\affiliation{Department of Physics \& Astronomy, University College London, Gower Street, London, WC1E 6BT, UK}
\author{E.~Buckley-Geer}
\affiliation{Fermi National Accelerator Laboratory, P. O. Box 500, Batavia, IL 60510, USA}
\author{A.~Carnero~Rosell}
\affiliation{Centro de Investigaciones Energ\'eticas, Medioambientales y Tecnol\'ogicas (CIEMAT), Madrid, Spain}
\affiliation{Laborat\'orio Interinstitucional de e-Astronomia - LIneA, Rua Gal. Jos\'e Cristino 77, Rio de Janeiro, RJ - 20921-400, Brazil}
\author{M.~Carrasco~Kind}
\affiliation{Department of Astronomy, University of Illinois at Urbana-Champaign, 1002 W. Green Street, Urbana, IL 61801, USA}
\affiliation{National Center for Supercomputing Applications, 1205 West Clark St., Urbana, IL 61801, USA}
\author{J.~Carretero}
\affiliation{Institut de F\'{\i}sica d'Altes Energies (IFAE), The Barcelona Institute of Science and Technology, Campus UAB, 08193 Bellaterra (Barcelona) Spain}
\author{F.~J.~Castander}
\affiliation{Institut d'Estudis Espacials de Catalunya (IEEC), 08034 Barcelona, Spain}
\affiliation{Institute of Space Sciences (ICE, CSIC),  Campus UAB, Carrer de Can Magrans, s/n,  08193 Barcelona, Spain}
\author{R.~Cawthon}
\affiliation{Physics Department, 2320 Chamberlin Hall, University of Wisconsin-Madison, 1150 University Avenue Madison, WI  53706-1390}
\author{L.~N.~da Costa}
\affiliation{Laborat\'orio Interinstitucional de e-Astronomia - LIneA, Rua Gal. Jos\'e Cristino 77, Rio de Janeiro, RJ - 20921-400, Brazil}
\affiliation{Observat\'orio Nacional, Rua Gal. Jos\'e Cristino 77, Rio de Janeiro, RJ - 20921-400, Brazil}
\author{J.~De~Vicente}
\affiliation{Centro de Investigaciones Energ\'eticas, Medioambientales y Tecnol\'ogicas (CIEMAT), Madrid, Spain}
\author{S.~Desai}
\affiliation{Department of Physics, IIT Hyderabad, Kandi, Telangana 502285, India}
\author{H.~T.~Diehl}
\affiliation{Fermi National Accelerator Laboratory, P. O. Box 500, Batavia, IL 60510, USA}
\author{J.~P.~Dietrich}
\affiliation{Excellence Cluster Origins, Boltzmannstr.\ 2, 85748 Garching, Germany}
\affiliation{Faculty of Physics, Ludwig-Maximilians-Universit\"at, Scheinerstr. 1, 81679 Munich, Germany}
\author{P.~Doel}
\affiliation{Department of Physics \& Astronomy, University College London, Gower Street, London, WC1E 6BT, UK}
\author{A.~E.~Evrard}
\affiliation{Department of Astronomy, University of Michigan, Ann Arbor, MI 48109, USA}
\affiliation{Department of Physics, University of Michigan, Ann Arbor, MI 48109, USA}
\author{B.~Flaugher}
\affiliation{Fermi National Accelerator Laboratory, P. O. Box 500, Batavia, IL 60510, USA}
\author{P.~Fosalba}
\affiliation{Institut d'Estudis Espacials de Catalunya (IEEC), 08034 Barcelona, Spain}
\affiliation{Institute of Space Sciences (ICE, CSIC),  Campus UAB, Carrer de Can Magrans, s/n,  08193 Barcelona, Spain}
\author{J.~Frieman}
\affiliation{Fermi National Accelerator Laboratory, P. O. Box 500, Batavia, IL 60510, USA}
\affiliation{Kavli Institute for Cosmological Physics, University of Chicago, Chicago, IL 60637, USA}
\author{J.~Garc\'ia-Bellido}
\affiliation{Instituto de Fisica Teorica UAM/CSIC, Universidad Autonoma de Madrid, 28049 Madrid, Spain}
\author{D.~W.~Gerdes}
\affiliation{Department of Astronomy, University of Michigan, Ann Arbor, MI 48109, USA}
\affiliation{Department of Physics, University of Michigan, Ann Arbor, MI 48109, USA}
\author{T.~Giannantonio}
\affiliation{Institute of Astronomy, University of Cambridge, Madingley Road, Cambridge CB3 0HA, UK}
\affiliation{Kavli Institute for Cosmology, University of Cambridge, Madingley Road, Cambridge CB3 0HA, UK}
\author{R.~A.~Gruendl}
\affiliation{Department of Astronomy, University of Illinois at Urbana-Champaign, 1002 W. Green Street, Urbana, IL 61801, USA}
\affiliation{National Center for Supercomputing Applications, 1205 West Clark St., Urbana, IL 61801, USA}
\author{J.~Gschwend}
\affiliation{Laborat\'orio Interinstitucional de e-Astronomia - LIneA, Rua Gal. Jos\'e Cristino 77, Rio de Janeiro, RJ - 20921-400, Brazil}
\affiliation{Observat\'orio Nacional, Rua Gal. Jos\'e Cristino 77, Rio de Janeiro, RJ - 20921-400, Brazil}
\author{W.~G.~Hartley}
\affiliation{Department of Physics \& Astronomy, University College London, Gower Street, London, WC1E 6BT, UK}
\affiliation{Department of Physics, ETH Zurich, Wolfgang-Pauli-Strasse 16, CH-8093 Zurich, Switzerland}
\author{D.~L.~Hollowood}
\affiliation{Santa Cruz Institute for Particle Physics, Santa Cruz, CA 95064, USA}
\author{D.~J.~James}
\affiliation{Center for Astrophysics $\vert$ Harvard \& Smithsonian, 60 Garden Street, Cambridge, MA 02138, USA}
\author{E.~Krause}
\affiliation{Department of Astronomy/Steward Observatory, University of Arizona, 933 North Cherry Avenue, Tucson, AZ 85721-0065, USA}
\author{K.~Kuehn}
\affiliation{Australian Astronomical Optics, Macquarie University, North Ryde, NSW 2113, Australia}
\author{N.~Kuropatkin}
\affiliation{Fermi National Accelerator Laboratory, P. O. Box 500, Batavia, IL 60510, USA}
\author{M.~A.~G.~Maia}
\affiliation{Laborat\'orio Interinstitucional de e-Astronomia - LIneA, Rua Gal. Jos\'e Cristino 77, Rio de Janeiro, RJ - 20921-400, Brazil}
\affiliation{Observat\'orio Nacional, Rua Gal. Jos\'e Cristino 77, Rio de Janeiro, RJ - 20921-400, Brazil}
\author{J.~L.~Marshall}
\affiliation{George P. and Cynthia Woods Mitchell Institute for Fundamental Physics and Astronomy, and Department of Physics and Astronomy, Texas A\&M University, College Station, TX 77843,  USA}
\author{P.~Melchior}
\affiliation{Department of Astrophysical Sciences, Princeton University, Peyton Hall, Princeton, NJ 08544, USA}
\author{F.~Menanteau}
\affiliation{Department of Astronomy, University of Illinois at Urbana-Champaign, 1002 W. Green Street, Urbana, IL 61801, USA}
\affiliation{National Center for Supercomputing Applications, 1205 West Clark St., Urbana, IL 61801, USA}
\author{R.~Miquel}
\affiliation{Instituci\'o Catalana de Recerca i Estudis Avan\c{c}ats, E-08010 Barcelona, Spain}
\affiliation{Institut de F\'{\i}sica d'Altes Energies (IFAE), The Barcelona Institute of Science and Technology, Campus UAB, 08193 Bellaterra (Barcelona) Spain}
\author{A.~A.~Plazas}
\affiliation{Department of Astrophysical Sciences, Princeton University, Peyton Hall, Princeton, NJ 08544, USA}
\author{A.~Roodman}
\affiliation{Kavli Institute for Particle Astrophysics \& Cosmology, P. O. Box 2450, Stanford University, Stanford, CA 94305, USA}
\affiliation{SLAC National Accelerator Laboratory, Menlo Park, CA 94025, USA}
\author{E.~Sanchez}
\affiliation{Centro de Investigaciones Energ\'eticas, Medioambientales y Tecnol\'ogicas (CIEMAT), Madrid, Spain}
\author{S.~Serrano}
\affiliation{Institut d'Estudis Espacials de Catalunya (IEEC), 08034 Barcelona, Spain}
\affiliation{Institute of Space Sciences (ICE, CSIC),  Campus UAB, Carrer de Can Magrans, s/n,  08193 Barcelona, Spain}
\author{I.~Sevilla-Noarbe}
\affiliation{Centro de Investigaciones Energ\'eticas, Medioambientales y Tecnol\'ogicas (CIEMAT), Madrid, Spain}
\author{M.~Smith}
\affiliation{School of Physics and Astronomy, University of Southampton,  Southampton, SO17 1BJ, UK}
\author{M.~Soares-Santos}
\affiliation{Brandeis University, Physics Department, 415 South Street, Waltham MA 02453}
\author{F.~Sobreira}
\affiliation{Instituto de F\'isica Gleb Wataghin, Universidade Estadual de Campinas, 13083-859, Campinas, SP, Brazil}
\affiliation{Laborat\'orio Interinstitucional de e-Astronomia - LIneA, Rua Gal. Jos\'e Cristino 77, Rio de Janeiro, RJ - 20921-400, Brazil}
\author{E.~Suchyta}
\affiliation{Computer Science and Mathematics Division, Oak Ridge National Laboratory, Oak Ridge, TN 37831}
\author{M.~E.~C.~Swanson}
\affiliation{National Center for Supercomputing Applications, 1205 West Clark St., Urbana, IL 61801, USA}
\author{G.~Tarle}
\affiliation{Department of Physics, University of Michigan, Ann Arbor, MI 48109, USA}
\author{R.~H.~Wechsler}
\affiliation{Department of Physics, Stanford University, 382 Via Pueblo Mall, Stanford, CA 94305, USA}
\affiliation{Kavli Institute for Particle Astrophysics \& Cosmology, P. O. Box 2450, Stanford University, Stanford, CA 94305, USA}
\affiliation{SLAC National Accelerator Laboratory, Menlo Park, CA 94025, USA}

\collaboration{DES Collaboration}

\date{Last updated \today}

\label{firstpage}

\begin{abstract}
An understanding of astrophysical feedback is important for constraining models of galaxy formation and for extracting cosmological information from current and future weak lensing surveys.  The thermal Sunyaev-Zel'dovich effect, quantified via the Compton-$y$ parameter, is a powerful tool for studying feedback, because it directly probes the pressure of the hot, ionized gas residing in dark matter halos.  Cross-correlations between galaxies and maps of Compton-$y$ obtained from cosmic microwave background surveys are sensitive to the redshift evolution of the gas pressure, and its dependence on halo mass. In this work, we use galaxies identified in year one data from the Dark Energy Survey and Compton-$y$ maps constructed from {\it Planck} observations.  We find highly significant (roughly $12\sigma$) detections of galaxy-$y$ cross-correlation in multiple redshift bins.  By jointly fitting these measurements as well as measurements of galaxy clustering, we constrain the halo bias-weighted, gas pressure of the Universe as a function of redshift between $0.15 \lesssim z \lesssim 0.75$.  We compare these measurements to predictions from hydrodynamical simulations, allowing us to constrain the amount of thermal energy in the halo gas relative to that resulting from gravitational collapse. 
\end{abstract}

\preprint{FERMILAB-PUB-19-161-A-AE}

\maketitle

\section{Introduction}
\label{sec:intro}

The nonlinear collapse of structure at late times leads to the formation of gravitationally bound dark matter halos. These massive objects are reservoirs of hot gas, with virial temperatures as high as $T \sim 10^8\,{\rm K}$.  This gas can be studied via its thermal emission, which is typically peaked in x-ray bands \citep[for a review, see e.g.][]{Bohringer:2010}.  Another way to study the gas in halos is via the thermal Sunyaev-Zel'dovich (tSZ) effect \citep{SZ_1972}, caused by inverse Compton scattering of CMB photons with the hot gas.  This scattering process leads to a spectral distortion which is observable at millimeter wavelengths \citep[e.g.][]{Carlstrom:2002}.  

The amplitude of the tSZ effect in some direction on the sky is characterized by the Compton-$y$ parameter, which is related to an integral along the line of sight of the ionized gas pressure.  By measuring contributions to $y$ as a function of redshift, we effectively probe the evolution of the gas pressure over cosmic time.  For the most massive halos, the evolution of the gas pressure is expected to be dominated by gravitational physics.  Gas falling into these halos is shock heated to the virial temperature during infall into the cluster potential \citep{Evrard:1990}.   For lower mass halos, on the other hand, other mechanisms may deposit energy and/or momentum into the gas; these mechanisms are generically referred to as "feedback."  

An understanding of baryonic feedback is important for constraining models of galaxy formation \citep[for a recent review, see][]{Naab:2017}.   Furthermore, since feedback can redistribute mass around halos (e.g. via gas outflows), an understanding of these processes is necessary for extracting cosmological constraints from small-scale measurements of the matter power spectrum with e.g. weak lensing surveys \citep{Rudd:2008,vanDaalen:2011}.

Because $y$ is sensitive to the line-of-sight {\it integrated} gas pressure, measurements of $y$ alone (such as the $y$ power spectrum) cannot be used to to directly determine the redshift evolution of the gas pressure.  However, given some tracer of the matter density field which can be restricted to narrow redshift intervals, cross-correlations of this tracer with $y$ can be used to isolate contributions to $y$ from different redshifts.  We take the cross-correlation approach in this analysis.

By cross-correlating a sample of galaxies identified in data from the Dark Energy Survey (DES)  \citep{Flaugher:2015} with $y$ maps generated from {\it Planck} data \citep{Planck:tsz}, we measure the evolution of the gas pressure as a function of redshift.  As we discuss in \S\ref{sec:formalism}, our cross-correlation measurements are sensitive to a combination of the gas pressure and the amplitude of galaxy clustering.  To break this degeneracy, we perform a joint fit to measurements of the galaxy-$y$ cross-correlation and to galaxy-galaxy clustering to constrain both the redshift evolution of the galaxy bias, and the redshift evolution of a term depending on the average gas pressure in dark matter halos. 

Our analysis relies on the so-called \redmagic galaxy selection from DES.  The \redmagic algorithm yields a sample of galaxies whose photometric redshifts are well constrained \citep{DES_redMaGiC}.   We note that we do not attempt to model the halo-galaxy connection for the \redmagic galaxies.  Rather, we use these galaxies only as tracers of the density field for the purposes of isolating contributions to $y$ from different redshifts.  Consequently, we will restrict our measurements to the {\it two-halo} regime, for which the galaxy-$y$ cross-correlation can be modeled without dependence on the precise way that \redmagic galaxies populate halos \citep[for a review of the halo model see][]{Cooray:2002}. 

Several previous analyses have also considered the cross-correlation between galaxy catalogs and Compton-$y$ maps from {\it Planck} \citep{Vikram:2017, Hill:2018, Makiya:2018, Tanimura2019}.  \citet{Vikram:2017} (hereafter \citetalias{Vikram:2017}) correlated {\it Planck} $y$ maps with a sample of galaxy groups identified from Sloan Digital Sky Survey (SDSS) data by \citet{Yang:2007}.  Our analysis differs from that of \citetalias{Vikram:2017} in several important respects.  First, the galaxy sample used in this analysis is derived from DES data, and extends to significantly higher redshift  ($z \sim 0.7$) than considered by \citetalias{Vikram:2017} ($z \lesssim 0.2$).  Additionally, while \citetalias{Vikram:2017} divided their correlation measurements into bins of halo mass, we divide our measurements into bins of halo redshift.  The measurements presented here can be considered complementary to those of \citetalias{Vikram:2017} with regard to constraining feedback models. 

\citet{Hill:2018} used measurements and modeling similar to \citetalias{Vikram:2017} in order to extract constraints on the halo $Y$-$M$ relation, finding hints of departure from the predictions of self-similar models at low halo masses.  Our approach is similar to that of \citet{Hill:2018}, although we only fit measurements in the two-halo regime.  

\citet{Planck:LBG} correlated galaxies identified in SDSS data with {\it Planck} $y$ maps.  The galaxy catalog used by \citet{Planck:LBG} was restricted to "isolated" galaxies in order to probe the pressure profiles of individual small mass halos (although note the issues with this approach pointed out by \citet{LeBrun15}, \citet{Greco:2015} and \citet{Hill:2018}).  Several authors have also investigated related correlations between Compton-$y$ and weak lensing \citep{VanWaerbeke:2014, Hill2013} .

Recently, \citet{Tanimura2019} measured the correlation of luminous red galaxies (LRGs) with the {\it Planck} $y$ maps in order to study astrophysical feedback.  Our analysis differs from that of \citet{Tanimura2019} in two crucial aspects. First, we are only interested in the galaxy-$y$ cross-correlations in the two-halo regime, whereas \citet{Tanimura2019} analyzed the full $y$ profile around LRGs, including in the one-halo regime. Second, and more importantly, the quantity of interest in the present work, namely the bias weighted pressure of the universe, is not sensitive to the connection between the galaxies used for cross-correlations and the parent halo, nor to the properties of the galaxies.  The analysis of \citet{Tanimura2019} exhibits strong dependence on the connection between stellar mass and halo mass for their LRG sample. 

The structure of the paper is as follows.  In \S\ref{sec:formalism} we present our model for the galaxy-$y$ and galaxy-galaxy cross-correlation measurements; in \S\ref{sec:data} we describe the DES, {\it Planck} and simulation data sets used in our analysis; in \S\ref{sec:analysis} we describe our measurement and fitting procedure, and validate this procedure by applying it to simulations; in \S\ref{sec:results} we present the results of our analysis applied to the data.  We conclude in \S\ref{sec:discussion}.

\section{Formalism}
\label{sec:formalism}

We are interested in modeling both the galaxy-$y$ and galaxy-galaxy correlation functions to extract constraints on the redshift evolution of the gas pressure.  Our analysis will focus on the large-scale, two-halo regime in which the details of the galaxy-halo connection can be ignored.  The primary motivation for this choice is that in the two-halo regime, the galaxy-$y$ cross-correlation function is insensitive to the details of the galaxy-halo connection, significantly simplifying the analysis. 

We will assume a fixed $\Lambda$CDM cosmological model throughout, and will therefore suppress dependence on cosmological parameters.  When analyzing the data, we adopt a $\Lambda$CDM model with $h = 0.7$, $\Omega_{\rm m} = 0.28$, $\Omega_b=0.044$, $n_s = 0.965$ and $\sigma_8 = 0.8$.  Given the uncertainties on our measurement of the galaxy-$y$ cross-correlation, adopting instead the best-fit cosmology from e.g. \citet{Planck:2018cosmo} has a negligible impact on our main constraints.

\subsection{Model for galaxy-$y$ cross-correlation}

The observed temperature signal on the sky in the direction $\hat{n}$ and at frequency $\nu$ due to the tSZ effect can be written as
\begin{eqnarray}
\Delta T(\hat{n}, \nu) = T_{\rm CMB} y(\hat{n}) f(\nu) ,
\end{eqnarray}
where $T_{\rm CMB} = 2.73\,{\rm K}$ is the mean temperature of the CMB, and $y(\hat{n})$ is the Compton-$y$ parameter.  In the non-relativistic limit, we have \citep{SZ:1980}:
\begin{eqnarray}
f(x = h\nu/k_B T_{\rm CMB}) = x \frac{e^x + 1}{e^x - 1} - 4,
\end{eqnarray}
where $h$ is Planck's constant, and $k_B$ is the Boltzmann constant.  

The Compton-$y$ parameter is in turn given by  (suppressing the directional dependence):
\beqa
y = \frac{\sigma_T}{m_e c^2} \int_0^{\infty} dl \, P_e(l),
\eeqa 
where $P_e(l)$ is the electron gas pressure (which dominates the inverse Compton scattering process that gives rise to the tSZ effect) at line of sight distance $l$, $\sigma_T$ is the Thomson cross section, $m_e$ is the electron mass and $c$ is the speed of light.   For a fully ionized gas consisting of hydrogen and helium, the electron pressure, $P_e$, is related to the total thermal pressure, $P_{th}$, by:
\beqa \label{eq:batt1}
P_e = \left[\frac{4-2Y}{8-5Y}\right] P_{th},
\eeqa
where $Y$ is the primordial helium mass fraction.  We adopt $Y = 0.24$.

We denote the galaxy-$y$ cross-correlation with $\xi_{yg}(R)$.  This quantity represents the expectation value of $y$ at transverse comoving separation $R$ from the galaxies in excess of the cosmic mean.  We work in comoving coordinates because this choice preserves the size of a halo of constant mass as measured by a spherical overdensity radius as a function of redshift.  We will use $r$ to denote the 3D comoving separation between the halo center and a given point.

The halo-$y$ cross-correlation function for galaxies at redshift $z$ can be written as 
\beqa \label{eq:xi_yg_proj}
\xi_{yg}(R , z) = \frac{\sigma_T}{m_e c^2} \frac{1}{1+z} \int_0^{\infty} d\chi \, \xi_{Pg}\left(\sqrt{\chi^2 + R^2},z\right),
\eeqa
where $\chi$ is the comoving distance along the line of sight, and $\xi_{Pg}(r,z)$ is the 3D correlation function between the electron pressure and the galaxy sample of interest \citepalias{Vikram:2017}.

As functions of cluster-centric distance, halo mass, and halo redshift, we write the halo electron pressure profile and total density profile as $P_e(r,M,z)$ and $\rho(r,M,z)$.  It is convenient to work with Fourier transformed quantities, rather than the real space ones, which we represent with $u_P(k,M,z)$ and $u_m(k,M,z)$, respectively.  For $u_P$, for instance, we have
\beqa
u_P(k,M,z) \equiv \int_0^{\infty} dr \, 4 \pi r^2 \frac{{\rm sin}(kr)}{kr} P_e(r,M,z).
\label{eq:press_Fourier}
\eeqa
An analogous equation holds for $u_M$.

The galaxy-pressure cross-correlation function can be related to the galaxy-pressure cross-power spectrum via
\begin{eqnarray}
\xi_{Pg}(r,z) = \int_0^{\infty} \frac{dk}{2\pi^2} k^2 \frac{\sin (kr)}{kr}P_{Pg}(k,z),
\end{eqnarray}
where $k$ is the wavenumber, and $P_{Pg}(k,z)$ is the galaxy-pressure cross-power spectrum.  This power spectrum can be decomposed into contributions from the halo in which the galaxy resides (i.e. one-halo) and contributions from other halos (i.e. two-halo):
\begin{eqnarray}
P_{Pg}(k,z) = P_{Pg}^{\rm one-halo}(k,z) + P_{Pg}^{\rm two-halo}(k,z).
\end{eqnarray}
The one-halo part is given by: 
\begin{multline}
P^{\rm one-halo}_{Pg}(k,z) = \int dM \, \frac{dn}{dM}  \frac{N(M,z)}{\bar{n}(z)} u_m(k,M,z) u_P(k,M,z),
\end{multline}
where $u_m(k,M,z)$ and $u_P(k,M,z)$ are the Fourier transforms of the halo mass and pressure profiles for halos of mass $M$ at redshift $z$. Here we have assumed that galaxies are distributed according to the dark matter profile. The average number of galaxies in a halo of mass $M$ at a redshift $z$ is given by $N(M,z)$ and the average number density of galaxies (across all masses) is given by $\bar{n}(z)$. The quantity $dn/dM$ is the halo mass function, specifying the number density of halos (per comoving volume) and per mass interval.  

The two-halo term is then:
\begin{multline}
P^{\rm two-halo}_{Pg}(k,M,z) =  \bigg[\frac{N(M,z)}{\bar{n}(z)} u_m(k,M,z)  \bigg] \times \\  
(1+z)^3 \bigg[\int dM' \bigg(\frac{dn}{dM'}\bigg)  u_P(k,M',z) P_{hh}(k,M,M')\bigg],
\end{multline}
where $P_{hh}$ is the halo-halo power spectrum.  In the two-halo limit, we can assume linear bias, i.e. $P_{hh}(k,M,M') = b(M)b(M')P_{\rm{lin}}(k)$.  
Note that the $(1+z)^3$ factor comes from converting between physical coordinates and comoving coordinates.

As stated above, we are interested here in the large scale, two-halo regime.  In this limit (i.e. $k \rightarrow 0$),
\begin{multline}
u_P(k\rightarrow0,M,z) = \int_0^{\infty} dr \, 4 \pi r^2 P_e(r,M,z) \equiv E_T(M,z),
\label{eq:press_avg}
\end{multline}
where we have defined $E_T$ as the total thermal energy in a halo of mass $M$ at redshift $z$.  Similarly, we have
\beqa
u_m(k\rightarrow0,M) = \int_0^{\infty} dr \, 4 \pi r^2 \frac{\rho(r,M)}{M} = \bigg\langle \frac{\rho}{M} \bigg\rangle.
\label{eq:mass_avg}
\eeqa

Consequently, in this limit, 
\begin{multline}\label{eq:Pgp2h}
P_{Pg}(k,z) = \bigg(\int_0^{\infty}b_g(M,z) \frac{dn}{dM} dM\bigg) \\ \bigg( (1+z)^3 \int_0^{\infty} dM' \frac{dn}{dM'} b(M',z) E_T(M',z) \bigg) P_{\rm lin}(k,z).
\end{multline}

\begin{figure}
	\centering
	\includegraphics[width=0.45\textwidth]{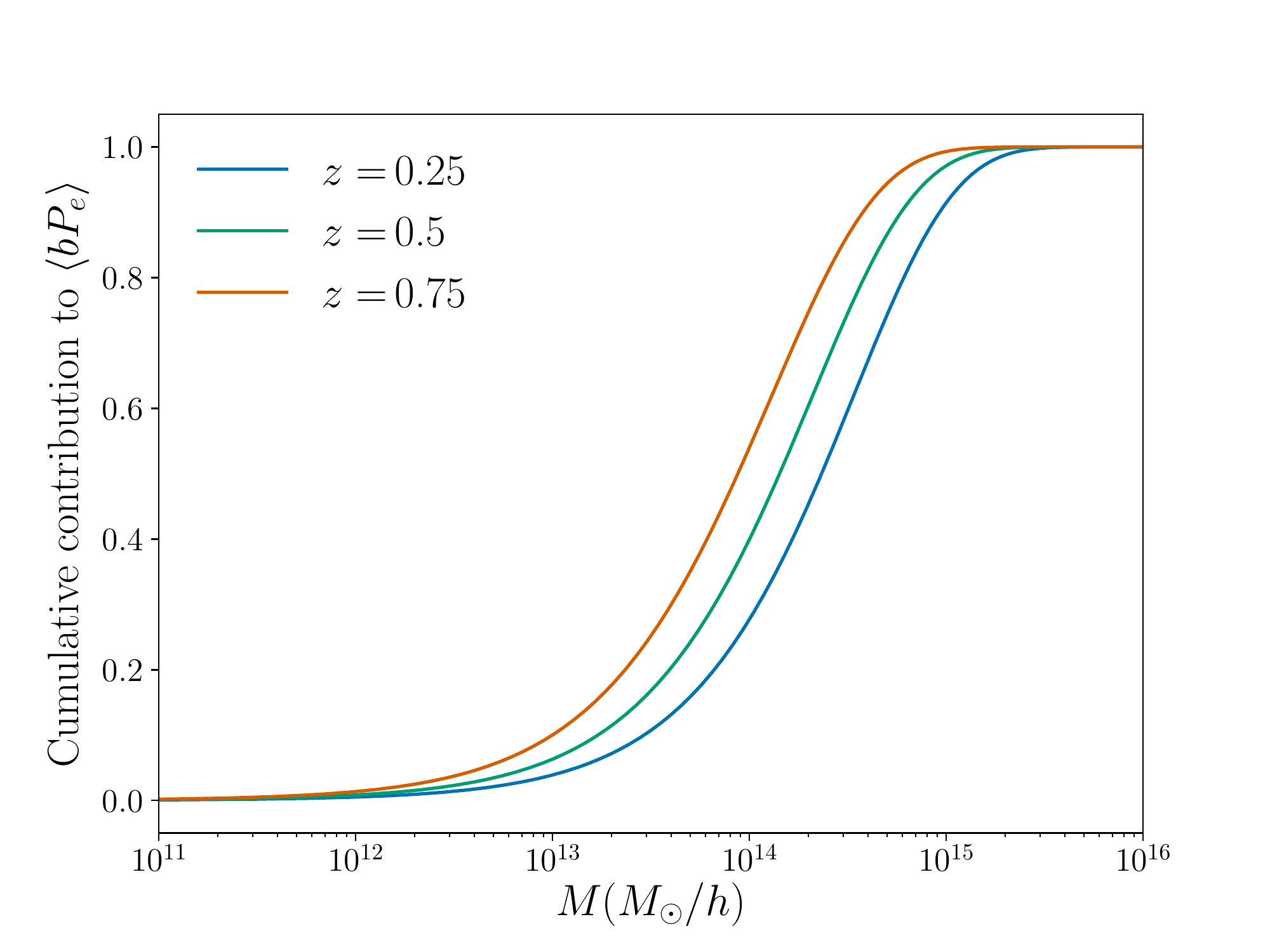}
	\caption{Cumulative contribution to the $\langle b P_e \rangle$ integral from theoretical estimates (using AGN feedback pressure profile described in \S\ref{sec:battaglia_profile}) of Eq.~\ref{eq:bpe} as a function of halo mass. Most contribution to the integral comes from halos in the range $10^{13}$ to $10^{15}$ $M_{\odot}/h$.  There is significant contribution to $\langle b P_e \rangle$ from halos with $M < 10^{14}\,M_{\odot}/h$; for current data, correlation analyses of the type considered here are the only way to probe this halo mass range.}
	\label{fig:bpe_integrand}
\end{figure}

We define the integral of $b_g$ over halos as the linear bias of our galaxy sample, i.e.
\beqa\label{eq:bg}
b_g(z) = \int_0^{\infty}\frac{N(M,z)}{\bar{n}(z)} \bigg\langle \frac{\rho}{M} \bigg\rangle \, b(M,z) \frac{dn}{dM} dM.
\eeqa
Eq.~\ref{eq:Pgp2h} can then be simplified further by defining:
\beqa\label{eq:bpe}
\langle b P_e\rangle (z) \equiv (1+z)^3 \int_0^{\infty} \frac{dn}{dM} b(M,z) E_T(M,z) dM.
\eeqa
This quantity represents the bias weighted thermal energy of all halos, and is the primary quantity of interest in this analysis. In order to estimate the $\langle b P_e \rangle$ from above equation, we use fitting formulae of halo mass function as described in \citet{Tinker:hmf} and large scale halo bias as descirbed in \citet{Tinker:bias}. We plot cumulative of the integrand of Eq.~\ref{eq:bpe} at several redshifts in Fig.~\ref{fig:bpe_integrand}.  The dominant contribution to $\langle b P_e \rangle$ comes from halos with masses in the range of about $3 \times 10^{12}   \lesssim M/(M_{\odot}/h) \lesssim 10^{15}$. 

In the two-halo limit, the galaxy-pressure cross-power spectrum then simplifies to:
\beqa
P^{\rm two-halo}_{Pg}(k,z) = b_g(z) \langle bP_e(z) \rangle P_{\rm lin}(k,z). 
\label{eq:two-halo_power_avg}
\eeqa
Substituting back into Eq.~\ref{eq:xi_yg_proj}, the two-halo contribution to the galaxy-$y$ cross-correlation function becomes
\begin{multline}
\xi^{\rm two-halo}_{yg}(R,z) = \frac{\sigma_T}{m_e c^2} b_g(z) \langle bP_e(z) \rangle  \\
\frac{1}{1+z} \int_{-\infty}^{\infty} d\chi \hspace{0.1cm} \xi_{\rm lin}\Bigg(\sqrt{\chi^2 + R^2},z \Bigg). 
\label{eq:yprof}
\end{multline}
The integral in the above equation is the projected linear correlation function, $w_{p,lin}(R)$. So, succinctly, our model for the cross-correlation function becomes:
\beqa
\xi^{\rm two-halo}_{yg}(R,z) = \frac{\sigma_T}{m_e c^2} b_g(z) \langle bP_e(z)\rangle \frac{w_{p,\rm{lin}}(R,z) }{1+z}.
\eeqa

A CMB experiment like {\it Planck} observes the sky convolved with a beam, which we must account for.  To do this, we first transform the above equation to angular space. Since $R$ denotes the comoving size of a halo, we have $\theta = R/\chi(z)$, where $\chi(z)$ is the comoving distance to redshift $z$. In Fourier space, the halo-y cross-power spectrum is: 
\begin{multline}
C^{\ell}_{yg} = \frac{\sigma_T}{m_e c^2} b_g(z) \langle bP_e \rangle  \int d\theta \, 2 \pi \theta J_0(\ell \theta)  \frac{w_{p,\rm{lin}}(\chi(z) \theta)}{1+z},
\end{multline}
where  $J_0$ is the Bessel function of the first kind. 

Multiplying this power spectrum by the beam function,  $B(\ell)$, and then inverse Fourier transforming, we obtain:
\beqa
\xi^{s,\rm two-halo}_{yg}(\chi\theta,z) = \int \frac{d\ell \, \ell}{2 \pi} J_0(\ell \theta) C_{yg}(\ell) B(\ell).
\label{eq:yprof_smooth}
\eeqa
We thus obtain in the two-halo limit (see also \citetalias{Vikram:2017}):
\beqa \label{gyz}
\xi^{s,\rm two-halo}_{yg}(R,z) \approx \frac{\sigma_T}{m_e c^2} b_g(z) \langle bP_e (z)\rangle \frac{w^S_{\rm lin}( R,z)}{1+z}, 
\eeqa
where $w^S_{\rm lin}(R,z)$ is the projected linear correlation function, smoothed by the beam as shown above.  

Eq.~\ref{gyz} describes the cross-correlation between galaxies and $y$ at a fixed redshift.  The \redmagic galaxies, however, are distributed over a broad redshift range, so we must average Eq.~\ref{gyz} over the normalized redshift distribution, $\omega^i(z)$, of the $i$th \redmagic galaxy bin.  Since the bias and bias-weighted pressure are expected to evolve slowly with redshift, and since the individual redshift bins of the \redmagic galaxies are only $\Delta z \sim 0.15$, we can define effective parameters over the whole bin, $ b_g $ and $ \langle bP_e \rangle $. The projected correlation function is also averaged across the redshift bins in this way.  Our final model for the galaxy-y cross-correlation is given by:
\begin{multline}
\label{eq:gyf}
\xi^{s,i}_{yg}(R >> r_{\rm vir},\bar{z}) \approx \frac{\sigma_T}{m_e c^2}  b^i_g \langle bP_e \rangle^i \int_0^{\infty}  \frac{w^S_{\rm lin}( R,z) \omega^i(z)}{1+z} dz.
\end{multline}
Given a cosmological model, $w^S_{\rm lin}( R)$ is fixed.  Consequently, specifying $b_g$ and $\langle b P_e \rangle$ is sufficient to specify the galaxy-$y$ cross-correlation function.  As we will show below, we can determine $b_g$ using fits to the galaxy-galaxy correlation function, allowing us to use the galaxy-$y$ measurements to solve for $\langle b P_e \rangle$.  

\subsection{Pressure profile model}
\label{sec:battaglia_profile}

Until now, we have been agnostic about the form of the halo pressure profile, $P_e(r,M,z)$.  \citet{Battaglia:2012} (hereafter \citetalias{Battaglia:2012}) measured the pressure profiles of halos in hydrodynamical simulations, and we will use fitting functions from those measurements in our analysis below.  The \citetalias{Battaglia:2012} fits use spherical overdensity definitions of the halo mass and radius, $M_{\Delta}$ and $R_{\Delta}$, respectively.  These are defined such that the mean density within $R_{\Delta}$ is $\Delta$ times critical density, $\rho_{\rm crit}(z)$, i.e.:
\begin{eqnarray}
M_{\Delta} = \Delta \frac{4}{3}\pi R_{\Delta}^3 \ \rho_{\rm crit}(z).
\end{eqnarray}
We will use both $\Delta = 200$ and $\Delta = 500$ definitions below where convenient.  The \citetalias{Battaglia:2012} pressure profile fitting function is then a generalized NFW model:
\begin{multline}
\label{eq:pfit}
P(x = r/R_{\Delta} , M_{\Delta}, z) = P_{\Delta} P_0 (x/x_c)^\gamma \left[1 + (x/x_c)^\alpha\right]^{-\beta},
\end{multline}
where $\gamma$, $\alpha$, $\beta$ and $x_c$ are redshift and mass dependent parameters of the model and the pressure normalization, $P_{\Delta}$, is given by:
\beqa
P_{\Delta} = \Delta \ \rho_{\rm crit}(z) \frac{\Omega_{b} }{\Omega_{m}} \frac{G M_{\Delta}}{2 R_{\Delta}},
\label{eq:pself_sim}
\eeqa
where $\Omega_b$ and  $\Omega_m$ are the baryon and matter fractions, respectively, at  redshift $z=0$.  Because of significant degeneracy between the parameters, \citetalias{Battaglia:2012} set $\alpha = 1.0$ and $\gamma = -0.3$.

The free parameters of the \citetalias{Battaglia:2012} fits are then $P_0$, $x_c$ and $\beta$.  \citetalias{Battaglia:2012} additionally modelled the mass and redshift dependence of these parameters using fits of the form
\begin{eqnarray}
A = A_0 \left( \frac{M_{200}}{10^{14} M_{\odot}}\right)^{\alpha_m} (1+z)^{\alpha_z},
\end{eqnarray}
where $A$ represents $P_0$, $x_c$ or $\beta$.  The best fit parameters are given in Table 1 of \citetalias{Battaglia:2012}.  

\citetalias{Battaglia:2012} considered different models for gas heating, described in more detail in \citet{Battaglia:2010} (hereafter \citetalias{Battaglia:2010}).  In our analysis of the data we primarily rely on the `shock heating' model from \citetalias{Battaglia:2010}.  In this model, gas is shock heated during infall into the cluster potential; no additional energy sources or cooling models are included.  Below, we will extend this model to include the possibility of additional energy sources, which we will use the data to constrain.  For the purposes of generating simulated $y$ maps, we will also employ the AGN feedback model from \citetalias{Battaglia:2010}, which includes a prescription for radiative cooling, star formation, and supernovae feedback, in addition to AGN.

The quantity $\langle b P_e \rangle$ depends on the full pressure profile of the halos, and is therefore sensitive to its behavior at large $r$. At distances $r \gtrsim 2 R_{200}$, \citetalias{Battaglia:2012} found that the pressure profile fits could depart from the mean profile in simulations by more than 5\%.  In our analysis, when computing $\langle b P_e \rangle$, we will truncate the model pressure profiles at $r = 3R_{500}$.  We will consider the impact of varying this choice in \S\ref{sec:bpe_constraints}.  Additionally, the $\langle b P_e \rangle$ integral receives some contribution from $M \sim 10^{13} \,M_{\odot}/h$ halos, below the halo mass limit of the \citetalias{Battaglia:2010} simulations.  Consequently, when we model $\langle b P_e \rangle$ we will effectively be extrapolating the \citetalias{Battaglia:2010} fits to a regime just below where they were calibrated.

\subsection{Model for additional energy sources}
\label{sec:additional_energy}

The main purpose of our analysis is to constrain the amount of  energy in the halo gas relative to that expected from gravitational collapse.  The energetics of the halo gas could be changed relative to the gravitational expectation by processes such as AGN feedback and cooling.  As described above, the observable quantity $\langle b P_e \rangle$ is sensitive to the total thermal energy in halos in the mass range from about $10^{13}$ to $10^{15}\,M_{\odot}$.  To constrain departures from the purely gravitational energy input to the gas, we adopt the model
\begin{eqnarray}
E_T(M) = E_T^{\rm sh}(M) (1 + \alpha(M)),
\end{eqnarray}
where $E_T^{\rm sh}(M)$ is the thermal energy computed as in Eq.~\ref{eq:press_avg} using the shock heating model for the pressure profile from \citetalias{Battaglia:2012} (i.e. gravitational energy input only, and no cooling).  We adopt a simple phenomenological model for $\alpha(M)$:
\begin{eqnarray}
\label{eq:energy_departure_model}
\alpha(M) = \begin{cases}
\alpha & \text{if\,}  M < M_{\rm th} \\
0 & \text{if\,} M > M_{\rm th} 
\end{cases},
\end{eqnarray}
where $\alpha$ is a constant.  The motivation for introducing $M_{\rm th}$ is that for very massive halos, we expect the gravitational energy to dominate over all other energy sources.  Below, we will set $M_{\rm th} = 10^{14}\,M_{\odot}$, although we will also consider the impact of taking $M_{\rm th} \rightarrow \infty$.

We emphasize that $\langle b P_e \rangle$ is sensitive to the {\it total} thermal energy in halos.  Any process which changes the pressure profile, but does not change the total thermal energy content should not change $\langle b P_e \rangle$.  Such process might include, for instance, bulk motions of gas.  An additional point worth emphasizing is that the $\langle b P_e \rangle$ measurements for a particular redshift bin constrain the total thermal energy in the halos at that redshift.  This thermal energy could be impacted by heating or cooling at {\it higher} redshift.  For instance, AGN feedback at $z > 1$ could impact the measured $\langle b P_e \rangle$, provided that gas has not had sufficient time to cool by the redshift of observation.

\subsection{Model for galaxy-galaxy clustering}

At fixed cosmology, Eq.~\ref{gyz} shows that the galaxy-$y$ cross-correlation in the two-halo regime is completely determined once $ \langle bP_e \rangle $ and $ b_g $ are specified.  We can break the degeneracy between the two quantities using information from galaxy clustering, which is sensitive to $ b_g $, but not $\langle b P_e \rangle$.  By performing a joint fit to the galaxy-$y$ and galaxy-galaxy correlation functions, we can therefore constrain $\langle bP_e \rangle $ as a function of $z$. 

To constrain $b_g$ we rely on measurements of galaxy-galaxy clustering.  We now develop a model for this observable in the two-halo regime.  The power spectrum of the galaxies in the two-halo regime is given by:
\begin{multline}
P^{\rm two-halo}_{gg}(k,z) = \\
\bigg[\int dM \frac{dn}{dM}\frac{N(M,z)}{\bar{n}(z)} u_m(k,M,z) b(M,z) \bigg]^2 P_{\rm lin}(k,z).
\end{multline}
In the two-halo regime, we can take the low-$k$ limit for the dark matter halo profile $ u_m(k,M,z) $, yielding:
\begin{multline}
P^{\rm two-halo}_{gg}(k,z) = \\
\bigg[\int dM \frac{dn}{dM}\frac{N(M,z)}{\bar{n}(z)} \bigg \langle \frac{\rho}{M} \bigg \rangle b(M,z) \bigg]^2 P_{\rm lin}(k,z).
\end{multline}

Using the same definition of $ b_g $ as in Eq.~\ref{eq:bg}, we find the galaxy-galaxy power spectrum to be:
\beqa
P^{\rm two-halo}_{gg}(k,z) = b_g(z)^2 P_{\rm lin}(k,z).
\eeqa
The Limber approximation \citep{Limber:1953,Loverde:2008} can then be used to relate the 3D power spectrum to the harmonic-space power spectrum on the sky:
\beqa
C_{gg}(\ell) = \int d\chi \frac{q^2_g(z) }{\chi^2} P_{\rm lin}\left(\frac{\ell +1/2}{\chi},\chi \right),
\eeqa
where $q$ is the weight function given by:
 \beqa
q_g(z) = b_g \omega(z) \frac{dz}{d\chi}.
\eeqa

The angular correlation functions can then be related to the harmonic cross-spectra for any given redshift bin \textit{i} via:
\beqa
w^{ii}(\theta) = \sum_{\ell} \frac{2 \ell +1}{4\pi} P_{\ell}\left(\cos(\theta)\right)\,C_{gg}^{ii}(\ell)\,
\eeqa
 where $P_{\ell}\left(\cos(\theta)\right)$ is the Legendre polynomial of the $\ell$-th order.  We note that this model is equivalent to that employed in the \citet{DESy1} analysis, which uses the same galaxy clustering measurements as employed here.

\section{Data}
\label{sec:data}

\subsection{DES \redmagic catalog}
\label{sec:des-cat}
The primary goal of this analysis is to constrain the redshift evolution of the pressure of the Universe by measuring the correlation between galaxies and maps of the Compton-$y$ parameter.  To this end, we require a sample of galaxies that have well measured redshifts, and which can be detected out to large redshift.  An ideal catalog for this purpose is the
\redmagic catalog \citep{DESy1} derived from first year (Y1) DES observations.

The Dark Energy Survey is a 5.5 year survey of 5000 sq. deg. of the southern sky in five optical bands ({\it g}, {\it r}, {\it i}, {\it z}, and $Y$) to a depth of $r > 24$.  In this analysis, we use first Y1 data from DES covering approximately 1321 sq. deg. to roughly $r \sim 23$ \citep{Flaugher:2015, Abbot:2016}.

\redmagic galaxies are identified in DES data based on a fit to a red sequence template using the methods described in \citet{DES_redMaGiC}.  The photometric accuracy of the selection is high: $\sigma_{\rm rmg} = 0.0167(1+z)$.  For details of the validation of the \redmagic redshift estimates, see \citet{DES_redMaGiC} and \citet{Cawthon:2018}.  

Throughout this analysis, we use the same selection of galaxies and redshift binning as used in the analysis of \citet{DESy1}.  Using the same selection as in \citet{DESy1} is advantageous since systematic errors in the redshift estimates for this sample have been thoroughly studied in \citet{Cawthon:2018}, and the impact of observational systematics on \redmagic galaxy detection have been studied in \citet{Elvin-Poole:2018}.  

The Y1 \redmagic sample was divided into five redshift bins from $z = 0.15$ to $z = 0.9$.  The first three redshift bins use a luminosity cut of $L/L_{*} > 0.5$, while the fourth and fifth redshift bins use cuts of $L/L_* > 1.0$ and $L/L_* > 1.5$, respectively, where $L_*$ is computed using a Bruzual and Charlot model \citep{Bruzual:2003}, as described in \citet{DES_redMaGiC}.  Given the small number of galaxies in the fifth bin and the potential for higher contamination of the galaxy-$y$ cross-correlation measurements in that bin (see below), we restrict our analysis to the first four redshift bins.

\begin{figure}
	\centering
	\includegraphics[width=0.45\textwidth]{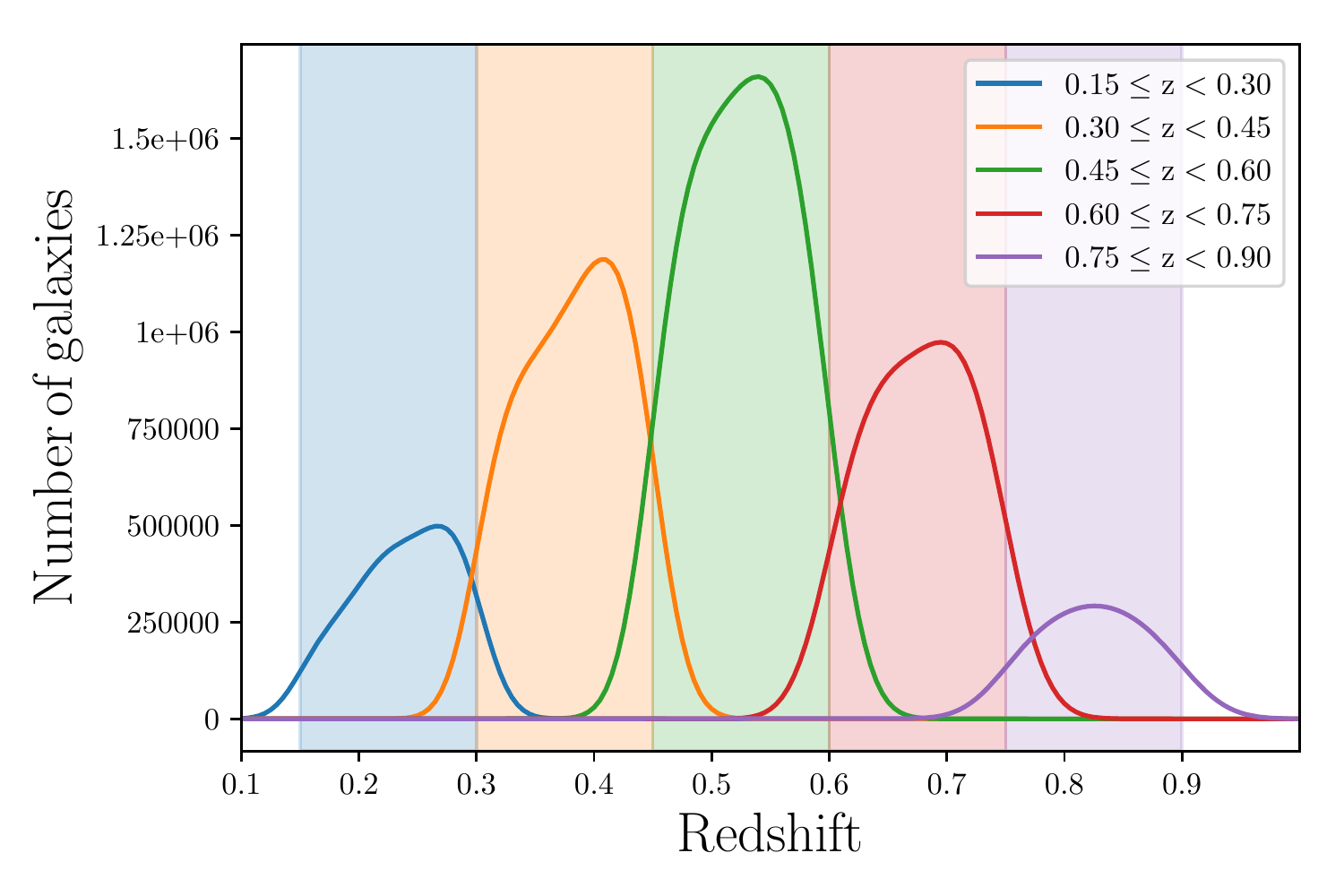}
	\caption{Redshift distributions of Y1 \redmagic galaxies used in this analysis. The galaxy sample is divided into five redshift bins, which are identical to the ones used in \citet{DESy1}. We only use the first four of these bins in the present analysis, as described in \S\ref{sec:des-cat}.  The integral of each curve over $dz$ is equal to the number of galaxies in the bin.  In total, the sample contains approximately 600,000 galaxies.	\label{fig:redmagic_z}
}
\end{figure}

Galaxies are placed into redshift bins based on their photometric redshift as estimated by the \redmagic algorithm \citet{DES_redMaGiC}.  \redmagic assigns a redshift estimate, $z_{\rm rmg}$, to each galaxy.  The estimated $\omega(z)$ for each bin is then computed as a sum of Gaussian probability distribution functions centered at $z^{i}_{\rm rmg}$, with standard deviation $\sigma_{\rm rmg}$.  The corresponding redshift distributions are shown in Fig.~\ref{fig:redmagic_z}.  

\subsection{{\it Planck} maps}
\label{sec:planck}

We correlate the \redmagic galaxies with maps of the Compton-$y$ parameter derived from {\it Planck} data.  {\it Planck} observed the sky in nine frequency bands from 30 GHz to 857 GHz from 2009 to 2013 \citep{Tauber2010,Planck2011}.  The resolution of the {\it Planck} experiment is band dependent, varying from roughly 30 arcminutes at the lowest frequencies to 5 arcminutes at the highest.

We use the publicly available 2015 {\it Planck} High Frequency Instrument (HFI) and Low Frequency Instrument (LFI) maps in this analysis \citep{Planck:lfi_maps,Planck:hfi_maps} and construct  Compton-$y$ maps using the Needlet Internal Linear Combination (NILC) algorithm that is described in \citet{Delabrouille:2009} and \citet{Guilloux:2007}. For comparison, we will also make use of the publicly available {\it Planck} estimates of $y$ described in \citet{Planck:tsz} which uses the same set of temperature maps. 

 While constructing various versions of Compton-$y$ map (see below), we use the same galactic mask as used in \citet{Planck:tsz} which blocks 2\% of the sky area (mostly in the galactic center). We also use the point source mask which is the union of the individual frequency point-source masks discussed in \citet{Planck:Point_source}. 

\subsection{Simulated sky maps}
\label{sec:sky_sims}

One of the primary concerns for the present analysis is possible contamination of the estimated $y$ maps by astrophysical foregrounds. The most significant potential contaminant is the cosmic infrared background (CIB), which is predominantly sourced by thermal emission from galaxies throughout the Universe.  CIB emission comes from a broad range of redshifts, roughly $z \sim 0.1$ to 4.0, with the bulk of emission coming from $z \gtrsim 1$ \citep[e.g.][]{Schmidt:2015}.  The majority of CIB emission is therefore beyond the redshift range of the galaxies considered in this analysis, and will therefore be uncorrelated with the \redmagic galaxies.  Such emission could constitute an additional noise source, but will not in general lead to a bias in the estimated galaxy-$y$ cross-correlation functions.  

However, some CIB emission is sourced from $z \lesssim 0.7$, which overlaps with the redshift range of the \redmagic galaxies.  Since the CIB is traces the large-scale structure, it will be correlated with the \redmagic galaxies.  Consequently, any leakage of CIB into the estimated $y$ maps over this redshift range could result in a bias to the estimated galaxy-$y$ cross-correlation functions.  

Another possible source of contamination is  bright radio sources.  Although the brightest sources are detected and masked, there will also be radio point sources that are not individually detected.   For instance, in a recent study by \citet{Shirasaki19}, it was found that radio sources can bias the tSZ-lensing correlation when using {\it Planck} data. Lastly, we may also have to worry about the potential biases and loss of signal-to-noise that may arise due to galactic dust contamination. We assess the effects of all the above mentioned biases using simulated sky maps as described below.  

We rely on both the  {\it Websky} mocks\footnote{\href{https://mocks.cita.utoronto.ca/index.php/Large_Scale_Structure_Mocks}{mocks.cita.utoronto.ca}} and the \citet{sehgal_2009} simulations. These two sets of simulations are useful in this analysis because they have produced correlated CIB maps and partially cover the frequency range used by {\it Planck}.  

The {\it Websky} mocks are full sky simulations of the extragalactic microwave sky generated using the mass-Peak Patch approach, which is a fully predictive initial-space algorithm, and a fast alternative to a full N-body simulation. As described in \citet{websky}, the mass-Peak Patch method finds an overcomplete set of just-collapsed structures through coarse-grained ellipsoidal dynamics and then resolves those structures further.  These maps are provided for frequencies 143, 217, 353, 545, and 857~GHz which are very similar to the {\it Planck} HFI channels. 

The \citet{sehgal_2009} simulations are another set of full sky simulations which provide maps for the cosmic microwave background, tSZ, kinetic SZ, populations of dusty star forming galaxies,  populations of galaxies that emit strongly at radio wavelengths, and dust from the Milky Way galaxy.  Maps are provided at six different frequencies: 30, 90, 148, 219, 277, and 350 GHz which are very similar to the {\it Planck} LFI channels and some of the HFI channels. These sets of maps allow us to directly test the effects of bright radio sources and galactic dust on the Compton-$y$ and its cross-correlation with halos that populate \redmagic-like galaxies.

We generate simulated sky maps in \texttt{Healpix}\footnote{\href{https://healpix.jpl.nasa.gov/}{healpix.jpl.nasa.gov}}format by combining the various component maps from the simulations described above.  For the {\it Websky} mocks, we combine Compton-$y$, lensed CMB and CIB; for the Sehgal simulations, we combine Compton-$y$, lensed CMB, CIB, radio galaxies and Milky Way galactic dust emission.  The "true" sky maps are then convolved with Gaussian beams with frequency-dependent full width half maxima (FWHM) corresponding to the {\it Planck} data.  Finally, we add {\it Planck}-like white noise to each channel at the levels given in Table 6 of \citet{Planck:hfi_response}.

\subsection{\mice and \buzzard N-body simulations}
\label{sec:nbody}

In addition to the estimation of $y$ from the {\it Planck} maps, the other major step in our analysis is the inference of $\langle b P_e \rangle$ from the measured correlation functions.  In order to test the methodology and assumptions involved in this step of the analysis, we rely on simulated \redmagic galaxy catalogs and $y$ maps.  The simulations used for this purpose are the \mice \citep{MICE:1,MICE:2,MICE:3} and \buzzard \citep{Buzzard:1} N-body simulations.  Both simulations have been populated with galaxy samples approximating \redmagic.

\mice  Grand Challenge simulation (MICE-GC) is an N-body simulation run on a 3 Gpc/$h$ box with $4096^3$ particles produced using the Gadget-2 code \citep{Gadget2}. The mass resolution of this simulation is $2.93 \times 10^{10} M_{\odot}/h$ across the full redshift range that we analyze here ($z < 0.75$), and halos are identified using a FoF algorithm using a linking length of 0.2. These halos are then populated with galaxies using a hybrid sub-halo abundance matching and a halo occupation distribution (HOD) approach, as detailed in \citet{MICE:2}. These methods are designed to match the joint distributions of luminosity, $g-r$ color, and clustering amplitude observed in SDSS \citep{SDSS:Zehavi_2011}.  The construction of the halo and galaxy catalogs is described in \citet{Crocce:2015}. A DES Y1-like catalog of galaxies with the spatial depth variations matching the real DES Y1 data is generated as described in \citet{DES:Bias}.  \mice assumes a flat $\Lambda$CDM cosmological model with  $h=0.7$, $\Omega_{\rm m}=0.25$, $\Omega_b=0.044$ and $\sigma_8=0.8$.  

\buzzard is a suite of simulated DES Y1-like galaxy catalogs constructed from dark matter-only N-body lightcones and including galaxies with DES $griz$ magnitudes with photometric errors, shape noise, and redshift uncertainties appropriate for the DES Y1 data \citep{Buzzard:1}.  This simulation is run using the code L-Gadget2 which is a proprietary version of the Gadget-2 code and the galaxy catalogs are built from the lightcone simulations using the ADDGALS algorithm \citep{Buzzard:1, DES:Bias, Weschler:inprep}.  Spherical-overdensity masses are assigned to all halos in \buzzard.  \buzzard assumes a flat $\Lambda$CDM cosmological model with  $h=0.7$, $\Omega_{\rm m}=0.286$, $\Omega_b=0.047$ and $\sigma_8=0.82$.

We generate mock Compton-$y$ maps for the N-body simulations by pasting $y$ profiles into mock sky maps at the locations of simulated halos.  The $y$ profile used for this purpose is the AGN feedback model (with $\Delta = 200$) from Table 1 of \citetalias{Battaglia:2012}. This approach to generating Compton-$y$ maps misses contributions to $y$ from halos below the resolution limit of the simulation.  However, given that \buzzard and \mice identify halos above $3 \times 10^{12} M_{\odot}/h$ and $10^{11} M_{\odot}/h$, respectively, Fig.~\ref{fig:bpe_integrand} shows that for both simulations, we capture at least 95\% of the contribution to $\langle b P_e \rangle$.  Since the statistical errors on the simulation measurements are significantly larger than $5\%$, any missing contribution to $\langle b P_e \rangle$ is not important for this work.  Note that since \mice uses only FoF masses, it is not strictly correct to apply the \citetalias{Battaglia:2012} profile to these halo mass estimates.  However, this inconsistency should not impact our validation tests described below.

\section{Analysis}
\label{sec:analysis}

\subsection{Measuring the galaxy-$y$ cross-correlation and galaxy-galaxy clustering}
\label{sec:measurement_procedure}

Our estimator for the galaxy-$y$ cross-correlation for galaxies in a single redshift bin and in the angular bin labeled by $\theta_{\alpha}$ is
\begin{eqnarray}
\hat{\xi}^{yg}(\theta_{\alpha}) = \frac{1}{N_D}\sum^{N_D}_{ij} y_m f(\theta_{ij}) - \frac{1}{N_R}\sum_{i_R j}^{N_R} y_m f(\theta_{i_R j}),
\end{eqnarray}
where $i$ ($i_R$) labels a galaxy (random point), $m$ labels a map pixel, $\theta_{im}$ is the angle between point $i$ and map pixel $m$, and $f$ is an indicator function such that $f(\theta) = 1$ if $\theta$ is in the bin $\theta_{\alpha}$ and $f(\theta) = 0$ otherwise.  The total number of galaxies and random points are $N_D$ and $N_R$, respectively.  By subtracting the cross-correlation of random points with $y$, we can undo the effects of chance correlations between the mask and the underlying $y$ field. 

We measure the galaxy-galaxy correlation using the standard \citet{Landy_Szalay93} estimator.  Because we use the same catalogs, redshift bins, and angular bins as in \citet{Elvin-Poole:2018}, our measurements of clustering of the \redmagic galaxies are identical to those in \citet{Elvin-Poole:2018}.  For both the galaxy-$y$ and galaxy-galaxy correlations, we compute the estimators using \texttt{TreeCorr} \citep{treecorr}.  

We measure the galaxy-$y$ cross-correlation in 20 radial bins from 1 Mpc/$h$ to 40 Mpc/$h$.  We measure galaxy-galaxy clustering in 20 angular bins from 2.5 arcmin to 250 arcmin which is the binning used in \citet{Elvin-Poole:2018}.  However, as described below in \S\ref{sec:model_fitting}, we do not include all measured scales when fitting these correlation functions, since the model is not expected to be valid at all scales.  Our angular scale cut choices are validated in \S\ref{sec:validation}.

\subsection{Covariance Estimation}
\label{sec:covariance}

Jointly fitting the measurements of the galaxy-$y$ and galaxy-galaxy correlations requires an estimate of the joint covariance between these two observables.  For this purpose, we use a hybrid covariance matrix estimate built from a combination of jackknife and theoretical estimates.  We validate the  covariance estimation in \S\ref{sec:validation}.

For the covariance block describing only the galaxy clustering measurements, we use the theoretical halo-model based covariance described in \citet{Krause:2017}.  This covariance has been extensively validated as part of the \citet{DESy1} analysis.

For the block describing the galaxy-$y$ covariance and for the cross-term blocks between galaxy-$y$ and galaxy clustering, we use jackknife estimates of the covariance.  The use of a jackknife is well motivated because several  noise sources in the $\hat{y}$ map are difficult to estimate.  These include noise from CIB and galactic dust.  Since the jackknife method uses the data itself to determine the covariance, it naturally captures these noise sources.  

The jackknife method for estimating the covariance of correlation functions on the sky is described in \citet{Norberg:2009}.  To construct jackknife patches on the sky, we use the KMeans algorithm\footnote{\href{https://github.com/esheldon/kmeans_radec}{https://github.com/esheldon/kmeans\_radec}}.  We find that 800 jackknife patches is sufficient for robust covariance estimation.  This means that each jackknife patch is approximately 85 arcmin across, which is approximately 1.5 times larger than our maximum measured scale for each redshift bin. 

Our jackknife estimates of the cross-covariance between the galaxy-clustering and galaxy-$y$ measurements are noisy.  When applying the jackknife covariance estimation to simulations (see \S\ref{sec:validation}), we find that this cross-covariance is largest when it is between two of the same redshift bins, as expected.  For the simulated measurements, zeroing cross-covariance between clustering and galaxy-$y$ measurements of {\it different} redshift bins has no impact on the inferred $\langle b P_e \rangle$.  To reduce the impact of noise in our covariance estimates, we therefore set these blocks to zero in our data estimate of the covariance.  The final covariance estimate is shown in Fig.~\ref{fig:cov_corr}.

\subsection{$y$ map estimation}
\label{sec:ymap}

\subsection{Overview}

The $y$ signal on the sky can be estimated as a linear combination of multi-frequency maps.  The constrained internal linear combination (CILC) method chooses weights in the linear combination that:
\begin{itemize}
    \item[] (a) impose the constraint that the estimator has unit response to a component with the frequency dependence of $y$, 
    \item[] (b) impose a constraint that the estimator has null response to some other component with known frequency dependence, 
    \item[] (c) minimize the variance of the estimator subject to the constraints from (a) and (b).  Below, we will consider several different analysis variations that attempt to null different components (or none at all).  
\end{itemize}

Note that the more components that are "nulled," the larger the variance of the resultant estimator, since imposing the nulling condition effectively reduces the number of degrees of freedom that can be used to minimize the variance.

When forming the estimated $y$ map with the CILC, the multi-frequency maps themselves must be decomposed into some set of basis functions, such as pixels or spherical harmonics.  In this analysis, we use maps decomposed using the needlet frame on the sphere \citep{Guilloux:2007,Marinucci:2008, Delabrouille:2009}.  The {\it Planck} estimate of $y$ generated using CILC methods in the needlet frame goes under the name Needlet Internal Linear Combination (NILC) and is described in \citet{Planck:tsz}.  We will use both the {\it Planck} NILC map and also construct our own versions for the purposes of testing biases due to contamination by the CIB and other astrophysical foregrounds.  A brief description of the analysis choices and methodology is given in  \S\ref{sec:yestimate_choices}; details are provided in  Appendix \ref{app:nilc}.

\subsubsection{Attempting to mitigate CIB bias in the $y$ map}\label{sec:yestimate_choices}

The {\it Planck} NILC $\hat{y}$ map \citep{Planck:tsz} enforces null response to components on the sky with the same frequency dependence as the CMB.  This choice is well motivated, since the CMB constitutes the dominant noise source over the frequency range that has significant signal-to-noise for the estimation of $y$.  We will refer to this choice as \texttt{unit-y-null-cmb}. We will also consider a variation that does not explicitly null any components, which we refer to as \texttt{unit-y}.

In the end, however, we only care about the cross-correlation of $\hat{y}$ with galaxies. The CMB correlates only very minimally with galaxies (due, for instance, to the integrated Sachs-Wolfe effect), and so should not result in a bias to the estimated galaxy-$y$ cross-correlation functions.  Since the CILC imposes a minimum variance condition on $\hat{y}$, explicitly nulling the CMB is not necessary for our purposes.  Attempting to null the CIB, on the other hand, is well motivated to prevent potential biases in the  $\langle bP_e \rangle$ estimation; we call this method \texttt{unit-y-null-cib}.  To null the CIB, one must adopt some reasonable choice for its frequency dependence.  Unfortunately, the frequency dependence of the CIB signal is uncertain, and furthermore, may vary with redshift, angular scale, or position on the sky.  

\begin{figure}
\centering
\includegraphics[width=1.0\linewidth]{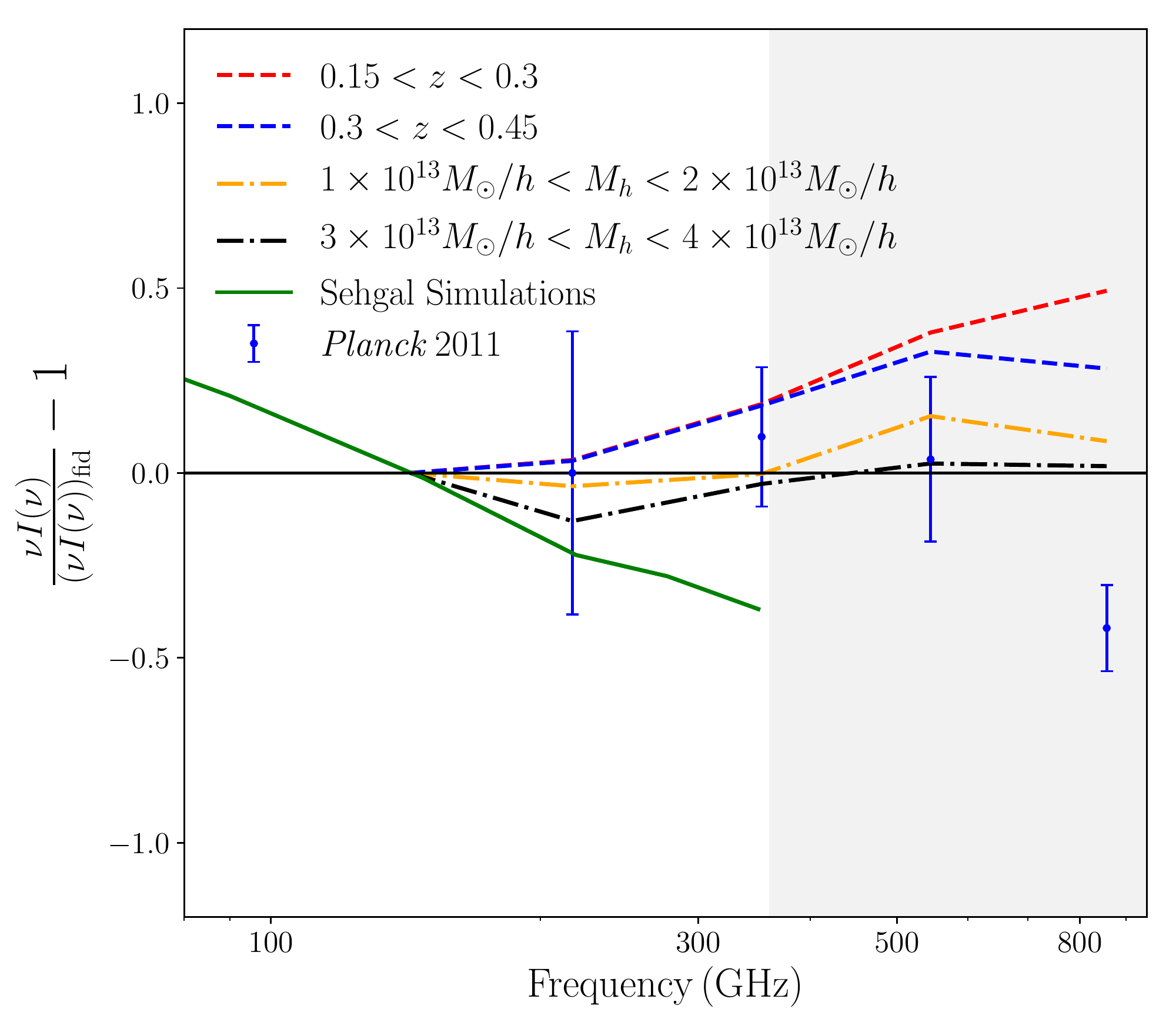}
	\caption{Frequency scaling of the halo-CIB correlation in the {\it Websky} mocks for different halo selections in redshift (dashed) and mass (dot-dashed).  Measurements are shown relative to the fiducial CIB model, as described in the text.  We also show the frequency scaling of the CIB in the Sehgal simulations (green solid curve), and the measurements from \citet{Planck2011} (blue points with errorbars). For frequencies less than 545~GHz (i.e. the frequency range used in this analysis, corresponding to the unshaded region in this plot), departures from our fiducial CIB model are less than 20\%, and are consistent with the {\it Planck} measurements. }
	\label{fig:cib-nu}
\end{figure}

We determine the frequency scaling of the CIB in the Sehgal simulations and the {\it Websky} mocks by cross-correlating the mock halos with the mock CIB maps.  To approximate the \redmagic selection, we correlate halos in the mass range $2 \times 10^{13} M_{\odot}/h < M < 3 \times 10^{13} M_{\odot}/h$ and  redshift range $0.45 < z < 0.6$ with the simulated CIB maps. We then measure the frequency scaling of these correlations at 100 arcmin, near the regime of interest for our $\langle b P_e \rangle$ constraints.  We compare this fiducial  CIB frequency dependence to {\it Planck} \citep{Planck:cib_scaling} and Sehgal simulations in Fig.~\ref{fig:cib-nu}.  The {\it Planck} points are derived from the rms fluctuations of the CIB anisotropy spectrum over the range $200 < \ell < 2000$.  We note these measurements are consistent with the frequency scaling of the mean of the CIB field, as described in \citet{Planck:cib_scaling}.  

Fig.~\ref{fig:cib-nu} shows that the frequency dependence of the CIB in both the simulations and the {\it Planck} data are consistent at roughly the 10\% level over the frequency range relevant to this analysis.  Larger deviations are observed at 545 and 857~GHz, but these channels are not used in the $y$ map reconstruction (see below). We also show the redshift dependence of the frequency scaling by cross-correlating with halos in different redshift bins, finding some variation. As mass of halos hosting the \redmagic galaxies is not completely certain, we also test the dependence of the CIB frequency scaling on the mass of halo used for cross-correlation.

The CIB intensity rises quickly at the higher frequency channels of {\it Planck}.  In order to reduce potential CIB contamination of the $y$ maps, we do not use the 545 or 857 GHz channels in our $y$ map reconstruction.  This choice differs from that made by \citet{Planck:tsz}, where both the 545 and 857 GHz channels were employed.  We see that variations in halo selection criteria impact the frequency dependence of CIB by less than 20\% for frequency channels below 545~GHz.  We have found that this choice makes the reconstructed $y$ maps less sensitive to the details of the CIB modelling, with only a minor degradation in signal-to-noise.

Finally, when analyzing the Sehgal mocks, we employ a large scale contiguous apodized mask that covers 10\% of the sky (near the galactic plane) in all the temperature maps to minimize the biases that might result from bright pixels in galactic plane. To minimize similar issues due to bright radio sources, we apply a point source mask that covers radio galaxies in the top decile. This mask is similar to the point source mask provided by the {\it Planck} collaboration that we use in the analysis of data. Since this is a highly non-contiguous mask, we inpaint masked pixels in the temperature maps. 

\subsubsection{Validation of $y$ estimation with mock skys}

\begin{figure}
\centering
\includegraphics[width=1.0\linewidth]{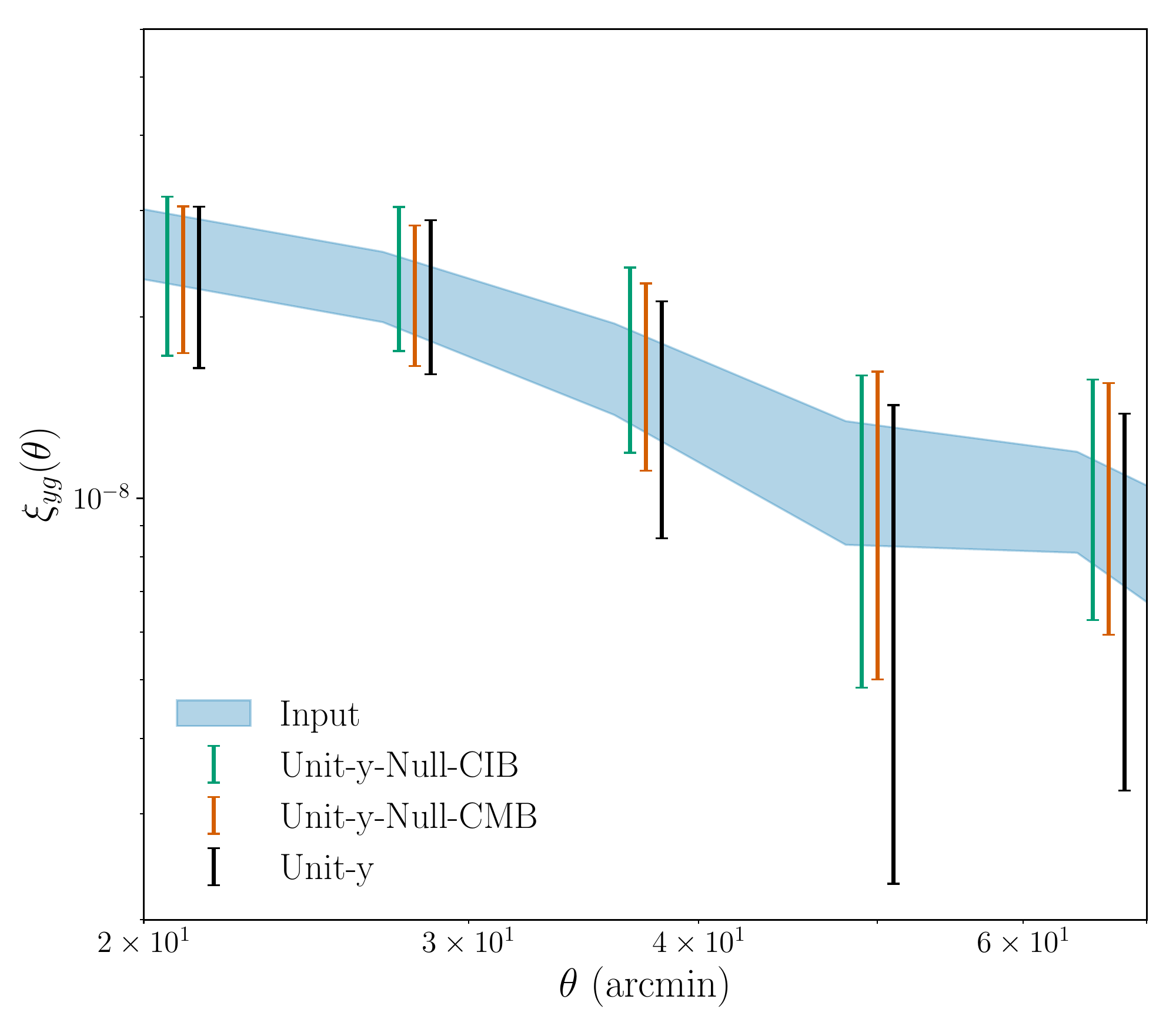}
	\caption{Galaxy-$y$ cross-correlation measurements with reconstructed $y$ maps from the Sehgal simulations.  We show results for the halo bin with $2\times 10^{13} M_{\odot}/h < M_h < 3 \times  10^{13} M_{\odot}/h$ and $0.15 < z < 0.3$.  The results for other redshift bins are similar.  We find that our $y$ reconstruction methods are sufficient to recover an essentially unbiased estimate of the halo-$y$ cross-correlation over the scales of interest. }
	\label{fig:compsep_sehgal}
\end{figure}

We apply our NILC pipeline to the simulated skies described in \S\ref{sec:sky_sims}, making the three nulling condition choices described above. We correlate the resultant $y$ maps with a sample of halos that approximate the \redmagic selection, with $2\times 10^{13} M_{\odot}/h < M_h < 3 \times  10^{13} M_{\odot}/h$.  The correlation results for the Sehgal simulation with halos in the redshift range $0.15 < z < 0.3$ are shown in Fig.~\ref{fig:compsep_sehgal}. In general, all three methods yield roughly consistent results that are also in good agreement with the true correlation signal.

The CIB model of the Sehgal simulations is not complete in the sense that it does not capture CIB contributions from halos below the mass limit of the simulation.  The CIB frequency model assumed in the Sehgal simulations is also somewhat out of date, and does not match current {\it Planck} observations.  For these reasons, we additionally use the {\it Websky} mocks for testing potential CIB biases.  The {\it Websky} mocks employ a model for CIB contributions from halos below the mass limit of the simulation, and also shows better agreement with recent {\it Planck} constraints on the CIB frequency dependence.  However, because the {\it Websky} mocks do not include radio sources or galactic dust, we primarily rely on the Sehgal simulations for validation.  We discuss tests using the {\it Websky} mocks in \S\ref{app:cita_val}.

\begin{figure*}
\includegraphics[width=0.48\linewidth]{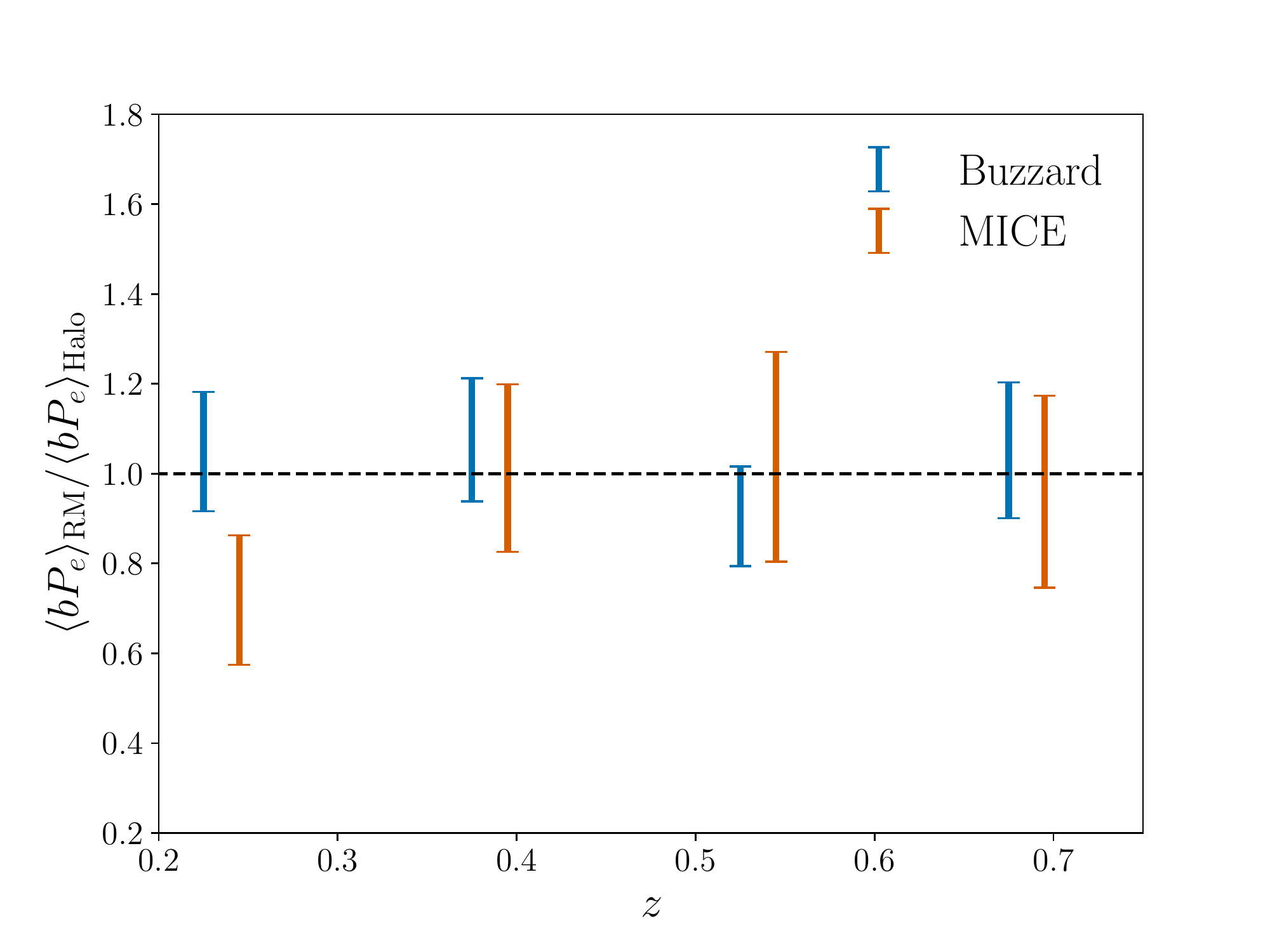}
\includegraphics[width=0.48\linewidth]{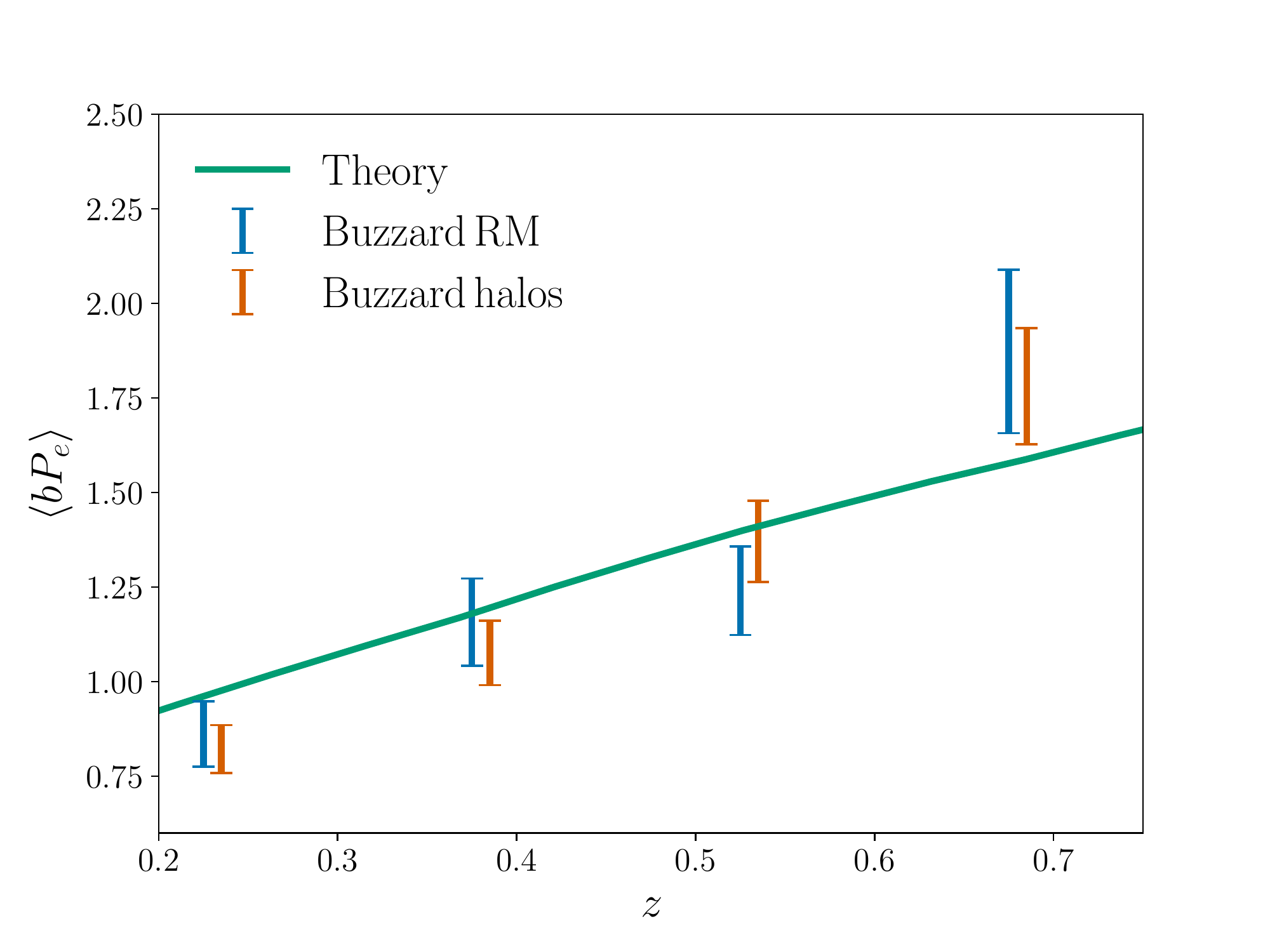}
\caption{Left panel shows the ratio of $\langle b P_e \rangle$ inferred in simulations from measurements with \redmagic galaxies to that inferred from halos.  For both the \buzzard (blue) and \mice (orange) simulations, the \redmagic galaxies and halos lead to consistent determinations of $\langle b P_e \rangle$.  This supports the notion that the measurements are sufficiently far in the two-halo regime that the inference of $\langle b P_e \rangle$ is independent of the halo-galaxy connection.  Right panel shows the measurements of $\langle b P_e \rangle$ in the \buzzard simulation compared to the theoretical prediction.  \label{fig:theory_check}}
\end{figure*}

\subsection{Model fitting}
\label{sec:model_fitting}

Our measurements of the galaxy-$y$ and galaxy-galaxy correlations in different redshift bins can be concatenated to form a single data vector
\begin{eqnarray}
\vec{d} = \left( d^{gg}_1, d^{gy}_1, d^{gg}_2, d^{gy}_2, \ldots , d^{gg}_4, d^{gy}_4 \right),
\end{eqnarray}
where $d^{gg}_i$ and $d^{gy}_i$ are the clustering and galaxy-$y$ correlations measurements in the $i$th redshift bin, respectively.  We consider a Gaussian likelihood for the data:
\begin{eqnarray}
\mathcal{L}(\vec{d} \ | \ \vec{\theta}) = -\frac{1}{2}\left( \vec{d} - \vec{m}(\vec{\theta}) \right)^T \mathbf{C}^{-1} \left( \vec{d} - \vec{m}(\vec{\theta}) \right),
\end{eqnarray}
where $\mathbf{C}$ is the covariance matrix described in \S\ref{sec:covariance}, $\vec{\theta}$ represents the model parameters (galaxy bias, $b_i$, and bias-weighted pressure, $\langle b P_e \rangle_i$ for redshift bin $i$) for all redshift bins, and $\vec{m}$ represent the model vector calculated as described in \S\ref{sec:formalism}. We adopt flat priors on all of the parameters, and sample the posterior using Monte Carlo Markov Chain methods as implemented in the code \texttt{emcee} \citep{emcee}.

We restrict our fits to the galaxy-galaxy correlation functions to scales $R > 8\,{\rm Mpc}/h$.
This restriction is imposed to ensure that the measurements are in the two-halo dominated regime, as discussed in \S\ref{sec:formalism}, and is consistent with the scale cut choices motivated in \citet{Krause:2017} and \citet{DES:Bias}.  

The determination of appropriate scale cuts for the galaxy-$y$ cross-correlation is somewhat more involved.  As described in Appendix~\ref{sec:ymap}, the Compton-$y$ map used in this analysis is smoothed with a beam of FWHM  of $10$ arcmin.  The beam has the effect of pushing power from small to large scales, and therefore shifts the location of the one-to-two-halo transition. For the highest redshift \redmagic bins, this shift can be significant and hence we have to increase our scale cuts as we go to higher redshift bins. For the bins detailed in \S\ref{sec:des-cat}, we ensure that we only include the scale cuts that are approximately twice the beam size away for any given redshift bin in our analysis. This results in minimum scale cuts for each of the four redshift bins at 4, 6, 8 and 10 Mpc/$h$. For the maximum scale cut, we make sure that for each redshift bin, the size of an individual jackknife patch is approximately 1.5 times the maximum scale cut for that particular bin. To obtain a sufficiently low-noise estimate of the covariance matrix from the jackknifing procedure,  we need of order 800 jackknife patches. These considerations yield maximum scale cuts for each of the 4 bins of 11, 17, 25 and 30 Mpc/$h$.

\subsection{Validation of model assumptions and pipeline}
\label{sec:validation}

\begin{figure*}
\centering
\includegraphics[width=1.0\linewidth]{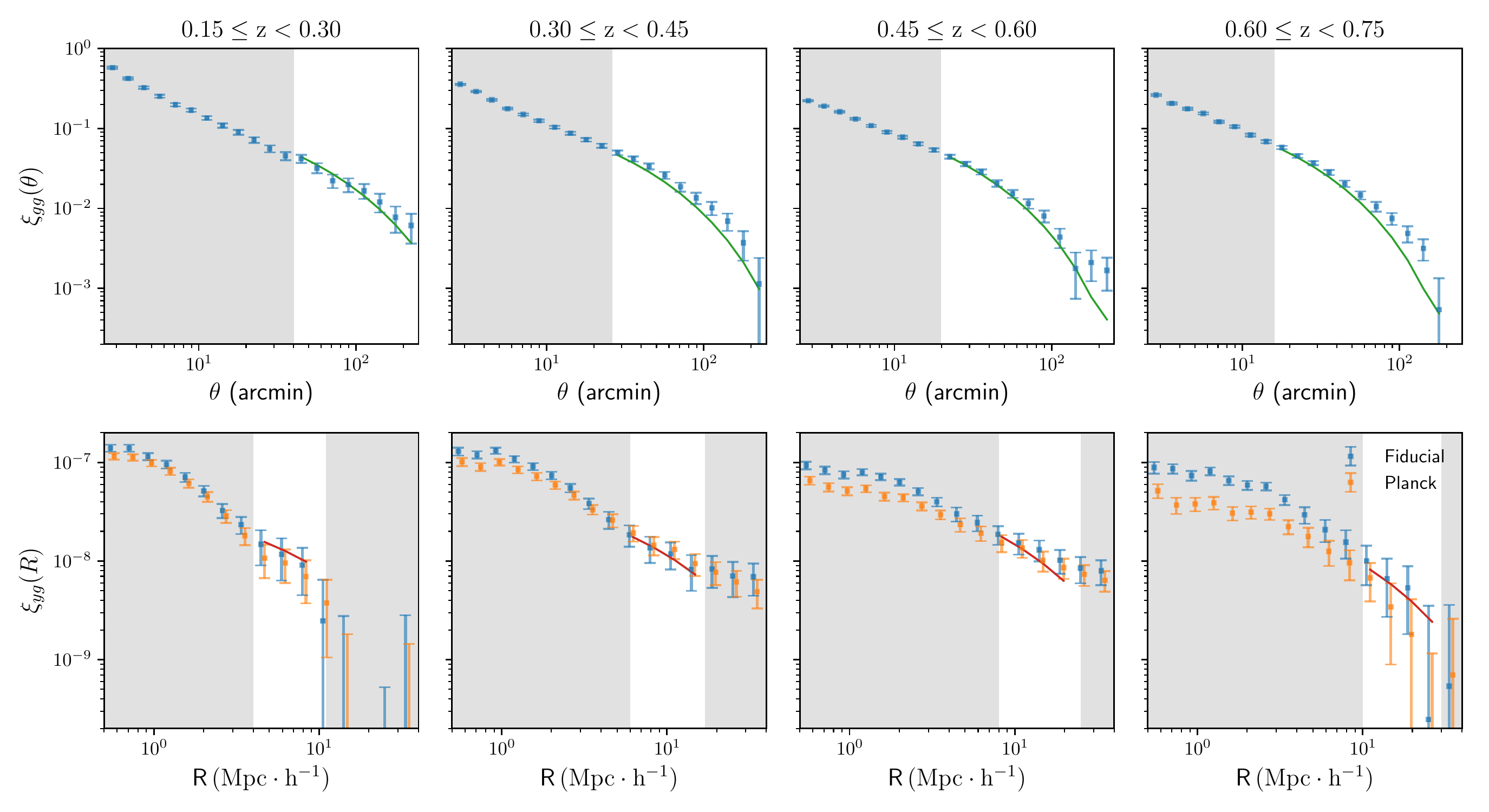}
\caption{Measurements of the galaxy auto-correlation (\textit{top row}) and Compton-y galaxy cross-correlation (\textit{bottom row}) at different redshift bins corresponding to four redshift bins used in the analysis. Solid line is the best-fit to the fiducial model of Compton-$y$ which is generated after removing 545GHz and 857GHz frequency channels from the analysis. Only data in the unshaded regions are used for fitting. These scale cut choices are validated in \S\ref{sec:validation}
}
\label{fig:corr_measurement}
\end{figure*}

\begin{figure}
\centering
\includegraphics[width=1.0\linewidth]{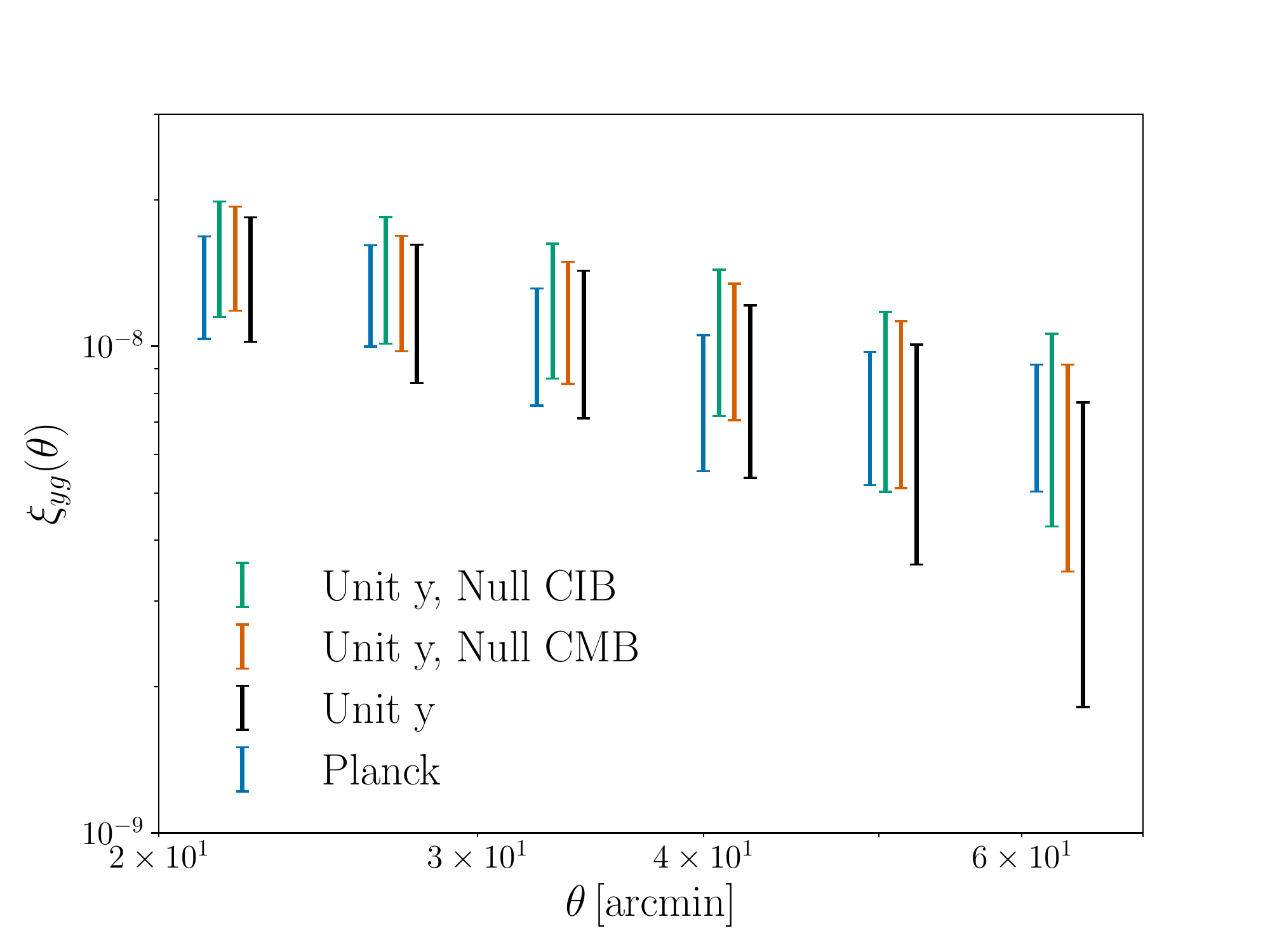}
	\caption{The galaxy-$y$ cross-correlation function over the scales of interest when the component separation method used to estimate $y$ is varied.  We show the correlation measurements for the highest signal to noise redshift bin, $0.45 < z < 0.6$, but results for the other redshift bins are similar.  We find that the estimated correlation function does not vary significantly when the $y$ estimation choices are varied.  Together with our validation with simulations, this constitutes strong evidence that our correlation measurements are not significantly biased by astrophysical contaminants in the estimated $y$.}
	\label{fig:compsep_dy_ac}
\end{figure}

We apply our analysis pipeline to the simulated data by correlating the mock $y$ maps with both the simulated \redmagic and halo catalogs.  In the two-halo regime, both the \redmagic galaxies and the halos should lead to consistent estimates of $\langle b P_e \rangle$.  The left panel of Fig.~\ref{fig:theory_check} shows the ratio of these two measurements for both the \buzzard and \mice simulations.  Indeed, we find that the \redmagic and halo measurements are consistent in both simulations, a strong test of our modeling assumptions and methodology.

We can also compare the recovered values of $\langle b P_e \rangle$ from the simulations to the value computed from the Eq.~\ref{eq:bpe}.  Since we know the true cosmological and profile parameters used to generate the simulated $y$ map, the measurement in simulations should match the theory calculation, provided our assumptions and methodology are correct.  The right panel of Fig.~\ref{fig:theory_check} shows this comparison (using both halos and \redmagic galaxies) for the \buzzard simulation. We find that the inferred values of $\langle bP_e \rangle$ are consistent with the theoretical expectation, providing a validation of our modeling, methodology, and scale cut choices.  Note that we do not perform this test with the \mice simulation, since as discussed in \S\ref{sec:nbody}, \mice uses FoF halo masses, while the \citetalias{Battaglia:2012} profile used to generate the simulated $y$ maps requires spherical overdensity masses.

\begin{figure*}
\includegraphics[width=0.85\linewidth]{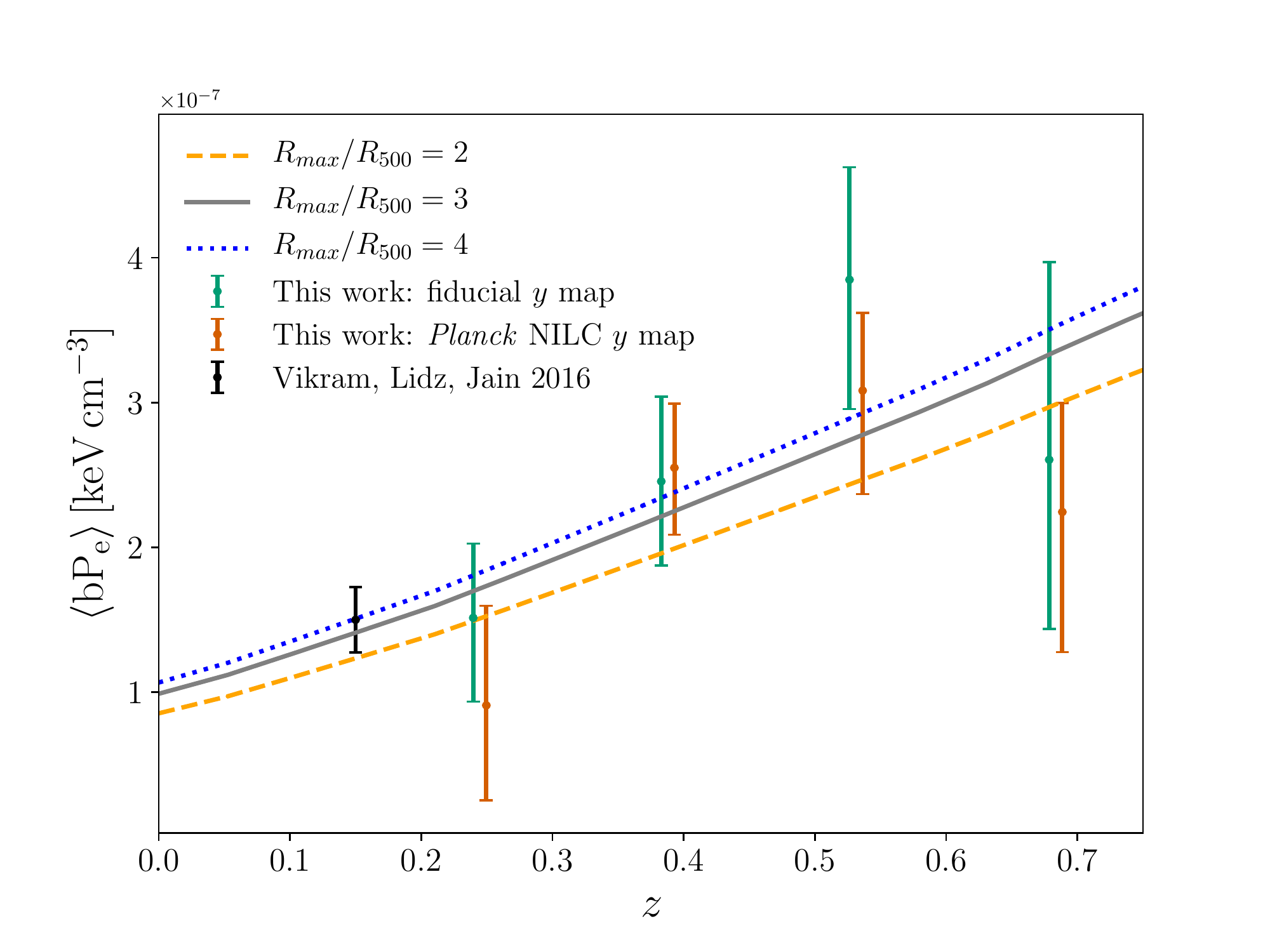}
\caption{Constraints on the redshift evolution of the bias weighted pressure of the Universe (Eq.~\ref{eq:bpe}). We compare the datapoints obtained from this work with \citetalias{Vikram:2017} and theory curves corresponding to shock heating model as described in \citetalias{Battaglia:2012}. For theory curves, all models are evaluated for $\Delta = 500$ and for various choices of $R_{max}/R_{\Delta}$.}.
\label{fig:bg_gp}
\end{figure*}

\begin{figure*}
\includegraphics[width=0.85\linewidth]{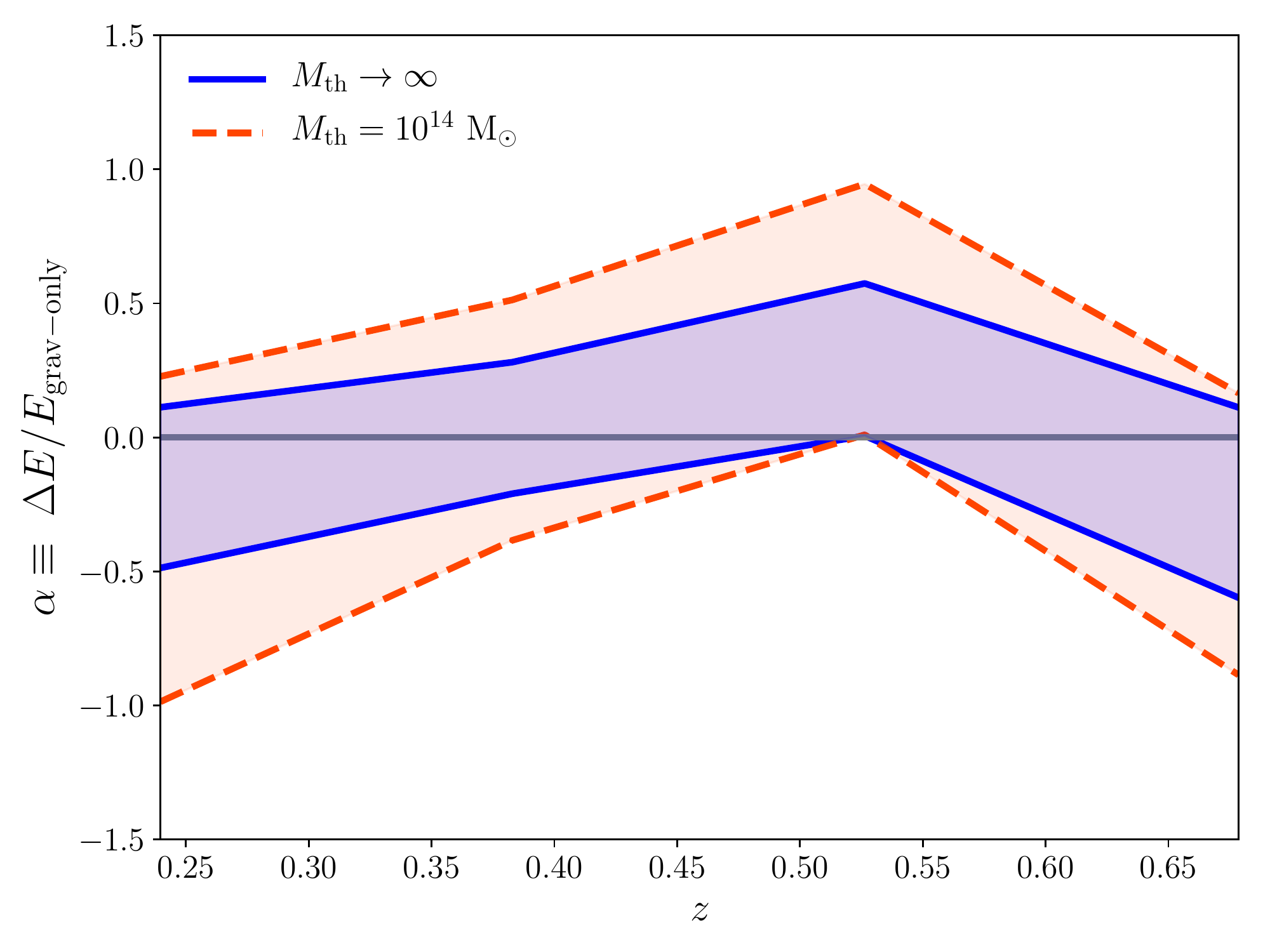}
\caption{Constraints on the thermal energy of the gas as a function of redshift.  The parameter, $\alpha$, defined in Eq.~\ref{eq:energy_departure_model}, measures the fractional departure of the gas thermal energy from the predictions of a model that only includes gravitational energy.  Large $\alpha$ means that some process must have contributed extra thermal energy to the gas, while negative $\alpha$ means that the gas must have cooled.  In our model, the addition (or deficit of) thermal energy impacts all halos below a threshold mass, $M_{\rm th}$.  We show the results for $M_{\rm th} \rightarrow \infty$ (region between blue solid curves) and for $M_{\rm th} = 10^{14}\,M_{\odot}$ (region between orange-red dashed curves).  The $\langle b P_e \rangle$ measurements presented in this work are sensitive to halos with $10^{13} \lesssim M \lesssim 10^{15}\,M_{\odot}$, as shown in Fig.~\ref{fig:bpe_integrand}.}
\label{fig:f_einj}
\end{figure*}

\section{Results}
\label{sec:results}

\subsection{Galaxy-$y$ cross-correlation measurements}

Our measurements of galaxy clustering (top) and the galaxy-$y$ correlation (bottom) using DES and {\it Planck} data are shown in Fig.~\ref{fig:corr_measurement}.  We show the galaxy-$y$ measurements with both our fiducial $\hat{y}$ map and the {\it Planck} $y$ map in Fig.~\ref{fig:corr_measurement}.  We obtain significant detections of galaxy-$y$ cross-correlation in all four redshift bins.  Across all radial scales, the galaxy-$y$ cross-correlation is detected at a significance of $12.3$, $12.9$, $12.2$ and $8.4\sigma$ for four redshift bins in order of increasing redshift. We restrict our model fits to the scales outside of the shaded regions to ensure that we remain in the two-halo regime where our modeling approximations are valid, as discussed in \S\ref{sec:analysis}.  The restrictions at large scales ensure that our jackknife estimate of the covariance is accurate; this cut leads to only a small degradation in signal-to-noise.

In order to assess potential biases in our measurements of the galaxy-$y$ cross-correlation, we repeat these measurements using the \texttt{unit-y-null-cib} and \texttt{unit-y} variations.  In the absence of a correlated contaminant in the estimated $y$ maps, different variations on the fiducial component separation choices should not lead to significant changes in the recovered mean galaxy-$y$ cross-correlation.  On the other hand, significant changes in the measured cross-correlation functions for varying component separation choices would be indicative of potential biases.  Note, though, that different component separation choices can lead to significant changes in the uncertainties on the estimates of the galaxy-$y$ cross-correlation, even in the absence of any contaminant.  

The impact of changing the component separation choices on the galaxy-$y$ cross-correlation measurements is shown in Fig.~\ref{fig:compsep_dy_ac}.  The results are shown only for the third redshift bin of \redmagic galaxies, since this has highest signal-to-noise.  The results obtained for the other redshift bins are similar. We find that the different $y$ estimation procedures yield statistically consistent measurements of the galaxy-$y$ cross-correlation over the range of scales used in this analysis.  These measurements are also consistent with the cross-correlations performed with the {\it Planck} $y$ map over the same range.  The insensitivity of the galaxy-$y$ cross-correlations to the component separation choices suggests that are our measurements are not biased by astrophysical contaminants.

However, as seen in Fig.~\ref{fig:corr_measurement}, there is a trend with increasing redshift for the {\it Planck} measurements at small scales to be lower in amplitude than the measurements with our fiducial $y$ map.  The main difference between our fiducial $y$ map and the {\it Planck} map is that we do not use the 545 and 857 GHz channels in our $y$ reconstruction, as described in \S\ref{sec:yestimate_choices}.  It is difficult to determine precisely the cause of the small scale discrepancy between the two $y$ map estimates seen in Fig.~\ref{fig:corr_measurement}.  It appears broadly consistent with contamination due to CIB, which would be expected to increase at higher redshift. We note that \citet{Planck:tsz} also found evidence for  CIB bias in the tSZ angular power spectrum at small scales.  We note, however, that the amount and direction of this CIB bias in the $y$ map obtained from NILC pipeline is sensitive to the frequency channels used, and that we consider here bias in galaxy-$y$ cross-correlation rather than the $y$ angular power spectrum considered in \citet{Planck:tsz}.  We emphasize, though, that over the range of scales fitted in this analysis, the estimates of the galaxy-$y$ correlations are consistent between the different $y$ maps.

\subsection{Constraints on bias-weighted pressure}\label{sec:bpe_constraints}

The quantity $\langle b P_e \rangle$, defined in Eq.~\ref{eq:bpe} represents the halo bias weighted thermal energy of the gas at redshift $z$. Fig.~\ref{fig:bg_gp} shows our constraints on this quantity as a function of redshift for two different $y$ maps: our fiducial \texttt{unit-y-null-cmb} map and the {\it Planck} NILC $y$ map.  The measurements with the two $y$ maps appear consistent, although precisely assessing the statistical consistency is complicated by the fact that the maps are highly correlated.  We find significant detections of $\langle b P_e \rangle$ in all redshift bins considered.  The multidimensional constraints on the model parameters are shown in Fig.~\ref{fig:full_posterior}.  

The black point in Fig.~\ref{fig:bg_gp} shows the constraint on $\langle b P_e \rangle$ from the analysis of \citetalias{Vikram:2017} using data from SDSS and {\it Planck}.  The \citetalias{Vikram:2017} point is at significantly lower redshift than the samples considered here ($z \sim 0.15$ as opposed to $0.2 \lesssim z \lesssim 0.75$).  The small errorbars on the \citetalias{Vikram:2017} measurements result from the large area of SDSS, roughly 10,000 sq. deg.  Our analysis with DES Y1 data uses roughly 1300 sq. deg, although the galaxy density of the DES Y1 measurements is significantly higher than the group catalog considered by \citetalias{Vikram:2017}.

\subsection{Constraints on feedback models}

The quantity $\langle b P_e \rangle$ depends on the cosmological parameters and on the pressure profiles of gas in halos.  Given the current uncertainty on the cosmological parameters from e.g. \citet{Planck:2018cosmo}, and the large model uncertainties on the gas profiles (especially at large radii), we focus on how $\langle b P_e \rangle$ can be used to constrain gas physics in this analysis.  Fig.~\ref{fig:bpe_integrand} shows that $\langle b P_e \rangle$ is sensitive primarily to halos with masses between $10^{13}$ and $10^{15}\,M_{\odot}$, with sensitivity to lower mass halos at high redshift.  Because $\langle b P_e \rangle$ effectively measures the total thermal energy in halos, it is particularly sensitive to the thermodynamics of gas in halo outskirts, where the volume is large.  As seen in \citetalias{Battaglia:2010}, it is precisely the large-radius, high-redshift regime probed in this analysis for which the predictions of different feedback models are significantly different.

The curves in Fig.~\ref{fig:bg_gp} show several predictions for the redshift evolution of $\langle bP_e \rangle$ for the `shock heating' model of \citetalias{Battaglia:2012} and \citetalias{Battaglia:2010}. In this model, the baryons are shock heated during infall into the cluster potential, and subsequently thermalize (with no AGN feedback or radiative cooling).  

We show several model predictions in Fig.~\ref{fig:bg_gp}, corresponding to different maximum radii for the halo gas profile.  In our fiducial analysis, we compute $\langle b P_e \rangle$ by integrating the pressure profile to $3R_{500}$. Similarly, the curve with $R_{\rm max}/R_{500} = a$ corresponds to integrating the profile to $a R_{500}$.  The data is consistent with shock heating models for $a = 2$, $a = 3$ and $a = 4$ with  $\chi^2/\rm d.o.f.$ of $2.9/4$, $2.11/4$ and $2.26/4$, respectively. 

For our fiducial shock heating model with $R_{\rm max} = 3 R_{500}$, we find $\chi^2$ per degree of freedom ($\rm d.o.f.$), $\chi^2/\rm d.o.f. = 2.11/4$ for the cross-correlation measurements with the \texttt{unit-y-null-cmb} map, and $\chi^2/\nu = 3.99/4$ for the cross-correlation with the {\it Planck} map.  In both cases, the data are statistically consistent with the shock heating model from \citetalias{Battaglia:2012}.

As described in \S\ref{sec:additional_energy}, the quantity $\langle b P_e \rangle$ is sensitive to the (bias weighted) total thermal energy in the halo gas.  We can use the measured $\langle  P_e \rangle $ to constrain any sources of energy beyond that associated with gravitational collapse, such as could be generated by feedback.  The additional energy model is described in \S\ref{sec:additional_energy}, and parameterizes any additional energy contributions for halos with mass $M < M_{\rm th}$ as a fractional excess, $\alpha(M)$, beyond that predicted by the shock heating model from \citetalias{Battaglia:2010}, which only includes gravitational energy.

The constraints on $\alpha(z)$ are shown in Fig.~\ref{fig:f_einj}.  In the limit that the threshold mass is very large ($M_{\rm th} \rightarrow \infty$, blue solid curve), we find that any mechanisms that change the thermal energy of the gas must not increase (or decrease) the thermal energy beyond about 30\% of the total gravitational energy over the redshift range $0.15 < z < 0.75$.  Note that this constraint applies to any thermal energy in the halos at that redshift.  If, for instance, significant energy injection occurred at higher redshift and the gas was not able to cool by redshift $z$, this injected energy would still contribute to our measurement.

The red dashed curve in Fig.~\ref{fig:f_einj} shows the impact of restricting the additional energy contributions to halos with $M < M_{\rm th} = 10^{14}\,M_{\odot}$.  The limit in this case is necessarily weaker since fewer halos contribute additional thermal energy. We find that over the redshift range probed and for halos with $M < 10^{14}\,M_{\odot}$, feedback (or other processes) must not contribute an amount of thermal energy greater than about 60\% of the halo gravitational energy (or reduce the thermal energy below about 60\% of the gravitational energy).  This constraint demonstrates part of the power of the $\langle b P_e \rangle$ constraints: we obtain constraints on additional energy input into low mass halos, even without explicitly probing the one-halo regime.

The implications of this constraint for feedback models depends, among other things, on how black holes populate their host halos and a careful comparison with simulations of AGN feedback is warranted. However, a rough estimate may nevertheless be helpful. A plausible estimate of the energy added by black hole feedback
is $E_{\rm feed} = \epsilon_r \eta M_{\rm BH} c^2$, where $\epsilon_r$ is the radiative efficiency and $\eta$ is the fraction of the radiated energy which couples (here thermally) to the surrounding gas.
Assuming $\epsilon_r=0.1$ and $\eta=0.05$ \citep{Matteo:05}, a black hole of mass $10^9 M_\odot$
adds $E_{\rm feed} = 9 \times 10^{60}$ ergs to the gas.  This is comparable to the thermal energy resulting from gravitational collapse (i.e. in the shock heating model) of a halo of mass $M_h=10^{13} M_\odot$, and 40\% of that of a $M_h=10^{14} M_\odot$ halo. This suggests that our constraints  --- limiting the extra thermal energy to about 60\% of the gravitational energy for halos with $M < M_h = 10^{14}\,M_{\odot}$ --- are reaching an interesting regime, and there are prospects to improve on them in the future.

It is also interesting to quantify the fraction of the total (i.e integrated over all redshifts) Compton-$y$ parameter accounted for in our measurements, which span roughly $z \sim 0.15$ to $z \sim 0.75$. Assuming the \citetalias{Battaglia:2012} shock heating pressure profile and $R_{\rm max} = 3 R_{500}$, the total average Compton-$y$ parameter is $\langle y \rangle = 2.9 \times 10^{-6}$ , while the contribution from the redshifts of the \redmagic sample, $0.15 \lesssim z \lesssim 0.75$ is $\langle y (0.15 \leq z \leq 0.75)\rangle = 6.7 \times 10^{-7}$ . In some sense, our measurement thererefore accounts for  $23\%$ of the total Compton-$y$ parameter (compared to only $2.5\%$ by the analysis of \citetalias{Vikram:2017}).  

One caveat to the above statements is that our analysis necessarily misses any unclustered contribution to the thermal energy.  Such a component would not be picked up in the galaxy-$y$ cross-correlation. Furthermore, we have not accounted for the possibility of overlapping halos in our halo model calculation.  If there is significant overlap of the pressure profiles, then we could be double counting some of hot gas.

\section{Conclusions}
\label{sec:discussion}

We have measured the cross-correlation of DES-identified galaxies with maps of the Compton-$y$ parameter generated from {\it Planck} data.  We detect significant cross-correlation in four redshift bins out to $z \sim 0.75$.  Using these measurements and measurements of galaxy clustering with the same galaxy sample, we constrain the redshift evolution of the bias-weighted thermal energy of the Universe, which we call $\langle b P_e \rangle$.  Our measurement of $\langle b P_e \rangle$ extends the previous measurement of this quantity from \citetalias{Vikram:2017} from $z \sim 0.15$ to $z \sim 0.75$.  High redshifts are of particular interest given the large uncertainties in both the modeling and data in this regime. 

Several features make $ \langle b P_e \rangle $ an interesting probe of gas physics.  First, it can be measured robustly even without a complete understanding of the galaxy-halo connection, as demonstrated in this analysis.  Second, $\langle b P_e \rangle$ is expected to be a sensitive probe of feedback models for several reasons.  First, unlike pressure profile measurements around massive clusters 
($M \gtrsim {\rm few} \times 10^{14}\,M_{\odot}$) (typically studied using x-ray measurements), the $\langle b P_e \rangle$ measurements probe mass scales down to $M \sim 10^{13}\, M_{\odot}/h$, and lower masses at high redshifts, as seen in Fig.~\ref{fig:bpe_integrand}.  It is precisely the low-mass halos for which feedback is expected to have a large impact.  Additionally, $\langle b P_e \rangle$ is sensitive to the outer pressure profiles ($R \gtrsim R_{\rm vir}$), as shown in Fig.~\ref{fig:bg_gp}.  As shown in \citetalias{Battaglia:2010}, various feedback prescriptions can make very different predictions in the outer halo regime.   Finally, $\langle b P_e \rangle$ probes the {\it total} thermal energy in halos.  Consequently, any process which changes the gas pressure profile, but does not inject or remove energy from the gas will not impact $\langle b P_e \rangle$.  For instance, our measurements would not be sensitive to feedback processes that only  move gas around without injecting any additional energy.   If one is interested in separating changes to the thermal energy from changes in the bulk distribution of gas, then  $\langle b P_e \rangle$ is a powerful tool to this end.

As shown in Fig.~\ref{fig:bg_gp}, our measurements are consistent with the shock heating model from \citetalias{Battaglia:2010}, with small variations depending on the extent of the profile.  We use the $\langle b P_e \rangle$ measurements to constrain departures from the purely gravitational shock heating model, with the results shown in Fig.~\ref{fig:f_einj}.  Our measurements constrain such departures at roughly the 20-60\% level.   

The measurements presented here use data from only the first year of DES observations, covering roughly 25\% of the full survey area of DES.  We also employ several conservative data cuts: (1) the highest redshift bin ($0.75 < z < 0.9$) is removed owing to low numbers of galaxies and greater potential for CIB contamination, (2) we restrict the measurements to only the two-halo regime, (3) we remove the largest angular scales due to the limitations of our jackknife covariance estimation.  With future improvements in data and methodology, these restrictions can be removed, enabling the full signal-to-noise of the measurements to be exploited. 

We also note that in the present analysis, we have assumed a fixed cosmological model.  This is reasonable given the uncertainties in our measurements and the precision of existing cosmological constraints.  However with future observations, it may be necessary to include uncertainty in cosmological parameters.

Current and future CMB observations will also enable higher signal-to-noise and higher resolution measurements of Compton-$y$.  Ground based CMB experiments like the South Pole Telescope \citep{Carlstrom:2011} and the Atacama Cosmology Telescope \citep{Swetz2011} have achieved significantly lower noise levels than {\it Planck} over significant fractions of the sky.  Ongoing CMB experiments like Advanced ACTPol \citep{Henderson:2016}, SPT-3G \citep{Benson:2014},  the Simons Observatory \citep{SimonsObs2018} and CMB Stage-4 \citep{Abazajian:2016} will yield very high signal-to-noise maps of $y$.  One challenge facing current and future ground based experiments, though, is potentially greater contamination of Compton-$y$ maps by foregrounds, owing to the narrower frequency coverage of these experiments.

The large apertures of ground based CMB experiments enables measurement of $y$ at significantly higher resolution than with {\it Planck}.  Because the analysis presented here was restricted to the two-halo regime, it is not necessarily the case that higher resolution measurements will dramatically extend the range of scales that can be exploited.  Some improvement is expected, though, especially for high-redshift galaxies, for which the beam pushes into the two-halo regime.  Future analyses with ground-based $y$ maps will gain significantly from using data in the one-halo regime. 

\section*{Acknowledgments}

We thank the {\it Websky} group for making their simulations publicly available, and for providing support for their use.  We thank Mathieu Remazeilles for assistance in understaning the {\it Planck} $y$-maps. We also thank Nicholas Battaglia for valuable discussions regarding pressure profile models and the formalism of the paper. 

SP, EB, ZX and JOS are partially supported by the U.S. National Science Foundation through award AST-1440226 for the ACT project. SP and EB are also partially supported by the US Department of Energy grant DE-SC0007901.

Funding for the DES Projects has been provided by the U.S. Department of Energy, the U.S. National Science Foundation, the Ministry of Science and Education of Spain, 
the Science and Technology Facilities Council of the United Kingdom, the Higher Education Funding Council for England, the National Center for Supercomputing 
Applications at the University of Illinois at Urbana-Champaign, the Kavli Institute of Cosmological Physics at the University of Chicago, 
the Center for Cosmology and Astro-Particle Physics at the Ohio State University,
the Mitchell Institute for Fundamental Physics and Astronomy at Texas A\&M University, Financiadora de Estudos e Projetos, 
Funda{\c c}{\~a}o Carlos Chagas Filho de Amparo {\`a} Pesquisa do Estado do Rio de Janeiro, Conselho Nacional de Desenvolvimento Cient{\'i}fico e Tecnol{\'o}gico and 
the Minist{\'e}rio da Ci{\^e}ncia, Tecnologia e Inova{\c c}{\~a}o, the Deutsche Forschungsgemeinschaft and the Collaborating Institutions in the Dark Energy Survey. 

The Collaborating Institutions are Argonne National Laboratory, the University of California at Santa Cruz, the University of Cambridge, Centro de Investigaciones Energ{\'e}ticas, 
Medioambientales y Tecnol{\'o}gicas-Madrid, the University of Chicago, University College London, the DES-Brazil Consortium, the University of Edinburgh, 
the Eidgen{\"o}ssische Technische Hochschule (ETH) Z{\"u}rich, 
Fermi National Accelerator Laboratory, the University of Illinois at Urbana-Champaign, the Institut de Ci{\`e}ncies de l'Espai (IEEC/CSIC), 
the Institut de F{\'i}sica d'Altes Energies, Lawrence Berkeley National Laboratory, the Ludwig-Maximilians Universit{\"a}t M{\"u}nchen and the associated Excellence Cluster Universe, 
the University of Michigan, the National Optical Astronomy Observatory, the University of Nottingham, The Ohio State University, the University of Pennsylvania, the University of Portsmouth, 
SLAC National Accelerator Laboratory, Stanford University, the University of Sussex, Texas A\&M University, and the OzDES Membership Consortium.

Based in part on observations at Cerro Tololo Inter-American Observatory, National Optical Astronomy Observatory, which is operated by the Association of 
Universities for Research in Astronomy (AURA) under a cooperative agreement with the National Science Foundation.

The DES data management system is supported by the National Science Foundation under Grant Numbers AST-1138766 and AST-1536171.
The DES participants from Spanish institutions are partially supported by MINECO under grants AYA2015-71825, ESP2015-66861, FPA2015-68048, SEV-2016-0588, SEV-2016-0597, and MDM-2015-0509, 
some of which include ERDF funds from the European Union. IFAE is partially funded by the CERCA program of the Generalitat de Catalunya.
Research leading to these results has received funding from the European Research
Council under the European Union's Seventh Framework Program (FP7/2007-2013) including ERC grant agreements 240672, 291329, and 306478.
We  acknowledge support from the Brazilian Instituto Nacional de Ci\^encia
e Tecnologia (INCT) e-Universe (CNPq grant 465376/2014-2).

This manuscript has been authored by Fermi Research Alliance, LLC under Contract No. DE-AC02-07CH11359 with the U.S. Department of Energy, Office of Science, Office of High Energy Physics. The United States Government retains and the publisher, by accepting the article for publication, acknowledges that the United States Government retains a non-exclusive, paid-up, irrevocable, world-wide license to publish or reproduce the published form of this manuscript, or allow others to do so, for United States Government purposes.

\clearpage 

\appendix

\section{NILC pipeline}\label{app:nilc}

In this appendix we elaborate on the $y$ map reconstruction pipeline. We follow the pipeline exactly as used in {\it Planck} $y$ map reconstruction with the freedom of changing the frequency dependence of the component that gets unit response as well as the number of components that get null response. The basic steps in the reconstruction are as follows:

\begin{enumerate}
	\item In the simulations, create the temperature maps by adding various relevant component \texttt{Healpix} maps of simulations at a given value of NSIDE. In the analysis using the {\it Websky} mocks and Sehgal simulations, we add the components described in \S\ref{sec:sky_sims} with NSIDE of 1024 and in common units of $\mu K_{CMB}$. In data we are given the temperature maps which we convert to common NSIDE of 1024 and to units of $\mu$ $K_{CMB}$ using the factors given in table 6 of \citet{Planck:hfi_response}
	
    \beqa
    T_{\nu}(\mathbf{\theta}) = a_{\nu}y(\mathbf{\theta}) + b^{(1)}_{\nu}C(\mathbf{\theta}) + b^{(2)}_{\nu}S(\mathbf{\theta}) + n_{\nu}(\mathbf{\theta}),
    \eeqa
    where $T_{\nu}(\mathbf{\theta})$ is the temperature map at a given frequency $\nu$ at $\mathbf{\theta}$ position in sky, $y(\mathbf{\theta})$ is the Compton-y map with $a_{\nu}$ frequency scaling, $C(\mathbf{\theta})$ is the CIB map (here we have assumed that it scales as $b^{(1)}_{\nu}$ across whole sky which may not be correct), $S(\mathbf{\theta})$ is the lensed CMB map and it scales as $b^{(2)}_{\nu}$ and $n_{\nu}(\mathbf{\theta})$ denotes all other components combined. For data, we download the publicly available temperature maps from the {\it Planck} collaboration \footnote{\href{https://pla.esac.esa.int/}{pla.esac.esa.int/}}. We also apply the relevant masks as described in the main text on these temperature maps before further processing .

	\item Smooth all the temperature maps ($T_{\nu} \rightarrow T_{\nu,s}$) with a Gaussian beam of FWHM = 10 arcmin. We choose this beam size as the Compton-y map by {\it Planck} Collaboration is also created with temperature maps smoothed with 10 arcmin beam.
	\beqa
	T_{\nu,s} = \mathcal{F}^{-1}(B(\ell)\times \mathcal{F}(T_{\nu})),
	\eeqa
	where $T_{\nu,s}$ are the smoothed temperature maps of frequency $\nu$ with gaussian window of given FWHM ($B(\ell)$). Here $\mathcal{F}$ denotes taking spherical harmonic transform to convert \texttt{Healpix} maps to $\ell,m$ space and $\mathcal{F}^{-1}$ takes the inverse fourier transform and converts back to map space. 
	\item Construct and save the spherical Fourier components, $T_{f,\nu}^{\ell,m}$ for each of above smoothed temperature maps ($f$ in the subscript denote the fourier space quantity).
	\item Use the 10 needlet band window functions ($h^i(\ell)$) provided by {\it Planck} Collaboration. These bands have the property that sum of square of all the bands is equal to 1 for all $\ell$. For each band, filter each frequency map with the corresponding window function. 
	\beqa
	\hat{T^{i}_{\nu}} = \mathcal{F}^{-1}(h^{i}(\ell)\times T_{f,\nu}^{\ell,m})
	\eeqa
	\item Calculate the weights for each frequency and needlet band corresponding to the input constraints for generating $y$ map. We always give unit response to Compton-$y$, that means we always have $\sum_\nu w_{\nu} a_{\nu} = 1$ for each needlet band $i$. Now, we experiment with either nulling one of the CIB signal and the CMB signal (nulling both would degrade our signal to noise) or not nulling any component. These weights are given by:
	\beqa
	\vec{w} = \frac{(\vec{b}^{(i),T} \mathbf{R}^{-1} \vec{b}^{(i)}) (\mathbf{R}^{-1} \vec{a}) - (\vec{b}^{(i),T} \mathbf{R}^{-1} \vec{a}) (\mathbf{R}^{-1} \vec{b}^{(i)}) }{(\vec{a}^T \mathbf{R}^{-1} \vec{a})(\vec{b}^{(i),T} \mathbf{R}^{-1} \vec{b}^{(i)}) - (\vec{a}^T \mathbf{R}^{-1} \vec{b}^{(i)})^2} ,
	\eeqa
	where $i$ can be 1 or 2 corresponding to the case of \texttt{unit-y-null-cib} and \texttt{unit-y-null-cmb} respectively. 
	Here $\mathbf{R}$ is the covariance caluclated in a smaller patch of sky that is determined by the maximum $\ell$ of each needlet band, number of frequencies and ilc-bias that we choose \citep{Delabrouille2014,Delabrouille:2009,Remazeilles2011,Remazeilles2013}. We choose an ilc bias ($b_{\rm ilc}$) value  of 0.1\%. This means that we need to calculate covariance using approximately ($\frac{N^i_{\nu}-1}{b_{\rm ilc}}$) pixels for any needlet band $i$, which uses $N^i_{\nu}$ channels for Compton-y estimation in any needlet band $i$.
	\item For each needlet band, $i$, multiply the weights obtained for each frequency with the needlet window filtered temperature maps. Now, sum all the resultant maps to get the final map for the given needlet band $i$.
	\item Now multiply the final map obtained for each band in previous step with the corresponding needlet window function and sum the resultant maps for all the bands. This gives us the estimated Compton-$y$ map for given sets of conditions and parameters. 
\end{enumerate}

\section{Validation of $y$ estimation on {\it Websky} mocks}
\label{app:cita_val}

As described in the text, the Sehgal CIB model is somewhat out of date, and is not expected to perfectly capture dependence of the CIB on frequency, redshift, and halo mass.  Consequently, we also test our $y$ estimation pipelines using the {\it Websky} mocks.

\begin{figure}
\centering
\includegraphics[width=1.0\linewidth]{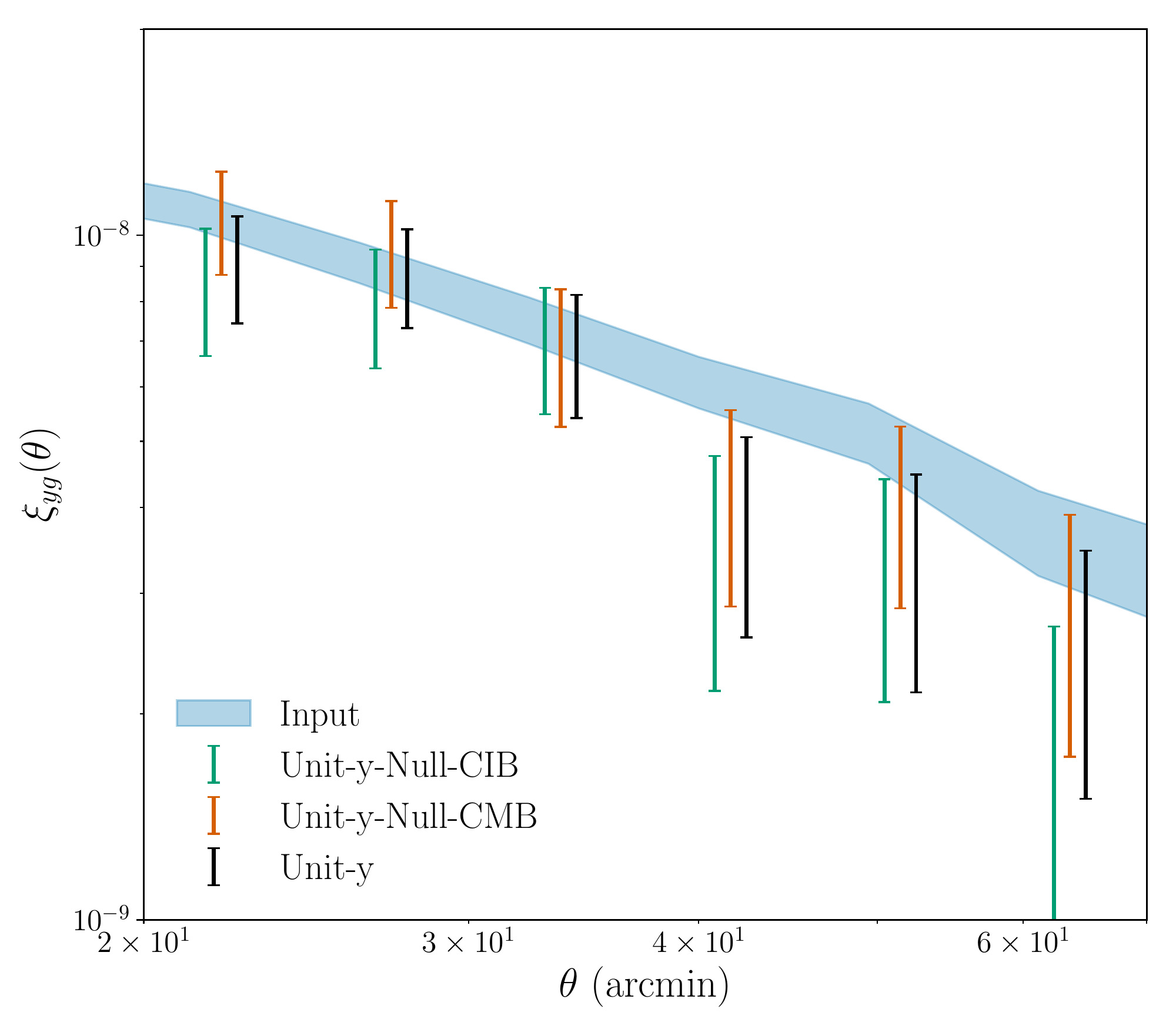}
	\caption{Cross-correlation of reconstructed Compton-$y$ with the halos in {\it Websky} mocks for various reconstruction methods. We correlate halos satisfying $0.3 < z < 0.45$ and $2\times 10^{13} M_{\odot}/h < M_h < 3 \times  10^{13} M_{\odot}/h$. The points labelled `input' correspond to the true halo-$y$ cross-correlation in the absence of any contamination.  The other points show the results of applying component separation to simulated sky maps that include the CIB signal.  In all cases, we use frequencies 100, 143, 217 and 343 GHz.  We find that the choice of unit-$y$, null-CMB leads to no significant bias in the inferred halo-$y$ cross-correlation.
	\label{fig:compsep_cita}}
\end{figure}

We reconstruct Compton-$y$ maps from the {\it Websky} mocks using the temperature maps corresponding to the frequencies less than 545GHz, as in our analysis of data.  We cross-correlate the reconstructed maps with halos in the mass range $2\times 10^{13} M_{\odot}/h < M_h < 3 \times  10^{13} M_{\odot}/h$. The result of this cross-correlation for the redshift bin $0.45 < z < 0.6$ is shown in Fig.~\ref{fig:compsep_cita}. We see that Compton-$y$ maps obtained from various choices of reconstruction methods, as detailed in \S\ref{sec:yestimate_choices}, result in halo-$y$ correlations that agree with each other as well as with the correlations with the true $y$ map. We find similar results for other redshift bins. As noted in the main text, since we do not have simulated radio galaxies for the {\it Websky} mocks, we rely mostly on the Sehgal simulations for validating our $y$ analysis choices.

\section{Covariance and Multidimensional Parameter constraints}

\begin{figure}[htbp]
\includegraphics[width=1.0\linewidth]{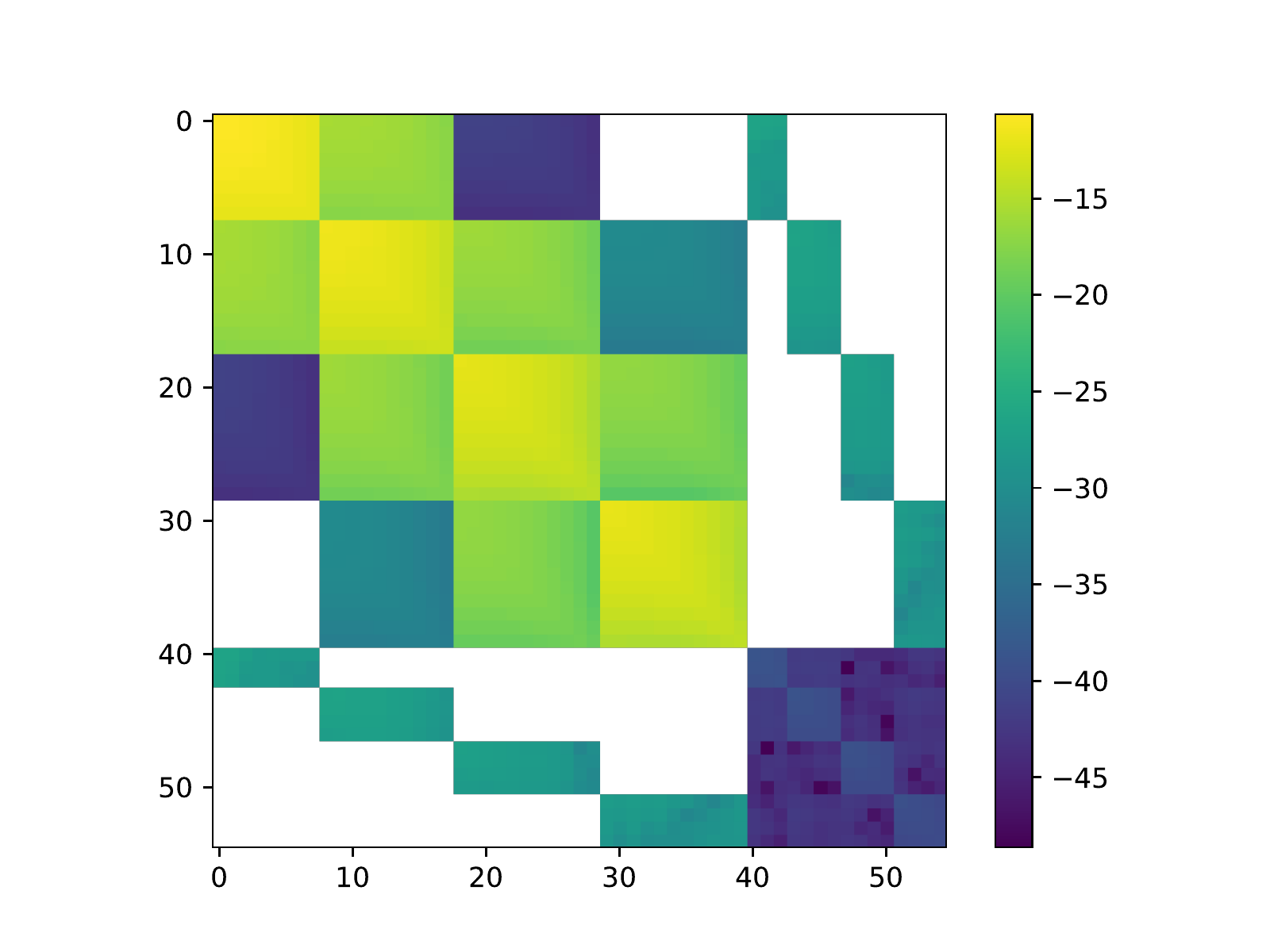}
\vfil
\includegraphics[width=1.0\linewidth]{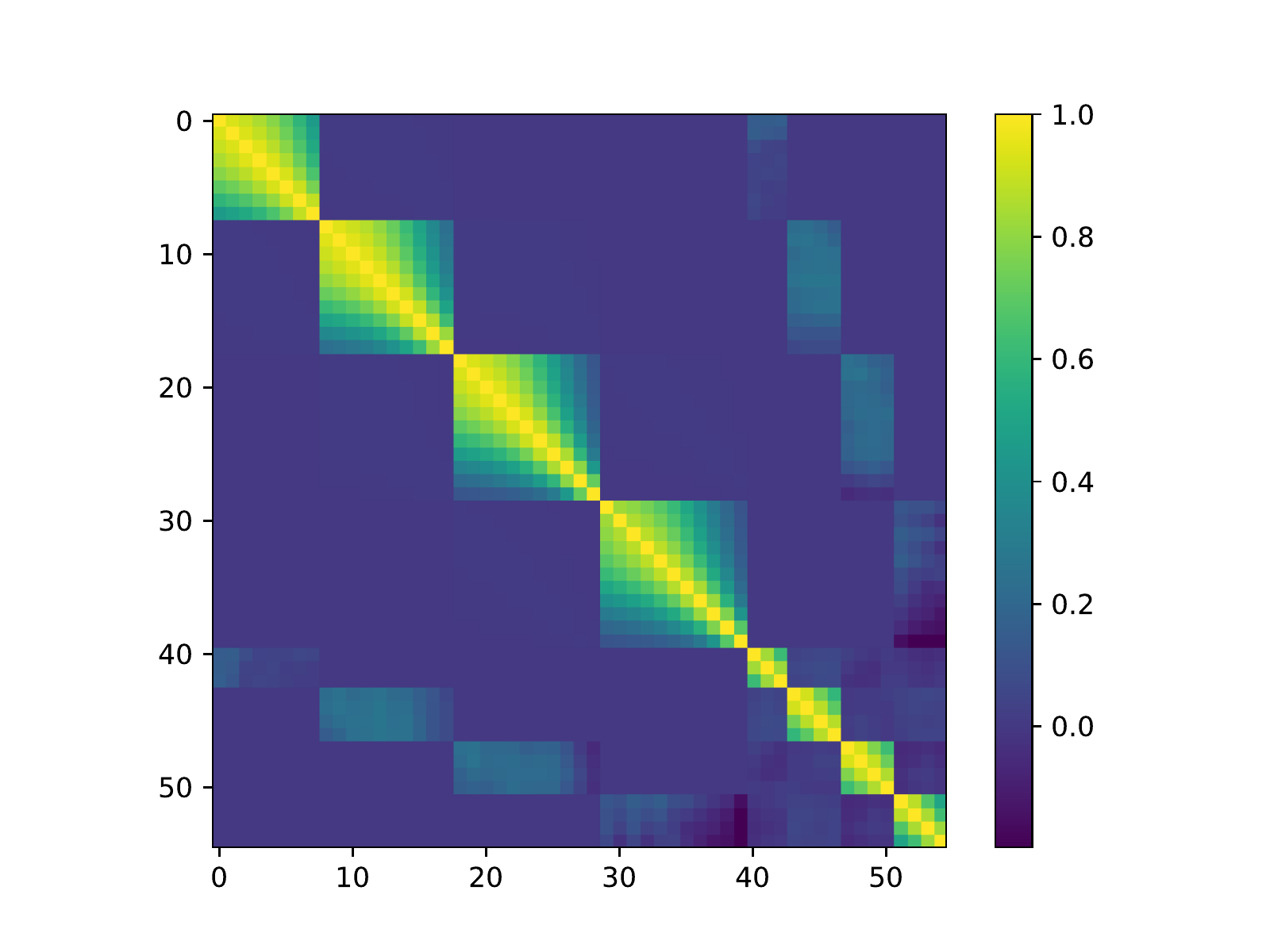}
\caption{Top panel shows the log of the absolute value of the final covariance matrix. Bottom panel shows the corresponding cross-correlation matrix } \label{fig:cov_corr}
\end{figure}

We show the estimated covariance and correlation matrices for the measurements in Fig.~\ref{fig:cov_corr}. As described in \S\ref{sec:covariance}, we use a jackknife resampling approach to estimating the blocks of the covariance matrix involving the galaxy-$y$ cross-correlation.  For the block involving only galaxy-galaxy clustering, we use the theoretical covariance estimate from \citet{Krause:2017}.  We also set to zero the cross redshift-bin covariance for the blocks corresponding to cross-covariance between galaxy-galaxy and galaxy-$y$.  

\begin{figure*}
\begin{center}
\includegraphics[width=0.8\linewidth]{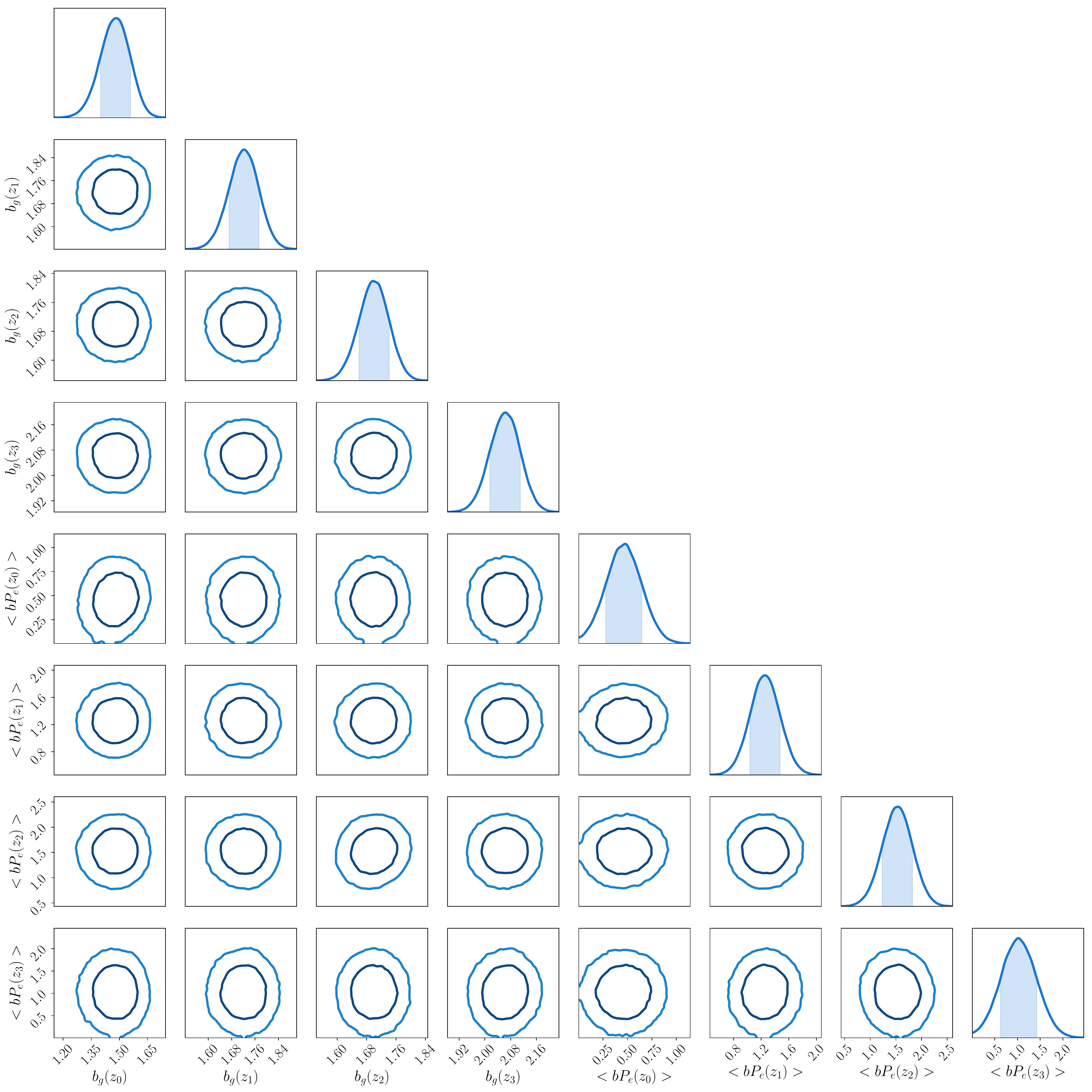}
\caption{Multi-dimensional parameter constraints from the model fits to data. First four parameters are galaxy bias for each of the four redshift bin used in this analysis and next four are bias weighted pressure corresponding to same bins}
\label{fig:full_posterior}
\end{center}
\end{figure*}

The multidimensional parameter constraints on the galaxy bias and $\langle b P_e \rangle$ parameters are shown in Fig.~\ref{fig:full_posterior} resulting from the MCMC analysis.  The MCMC is well converged, and there are no strong degeneracies between the parameters.

\clearpage

\bibliographystyle{mnras} 
\bibliography{ref}

\end{document}